\documentclass[11pt]{iopart}
\usepackage{hhline}
\usepackage{iopams}
\usepackage{amsfonts}
\usepackage{epsfig}
\usepackage{bbm}
\usepackage{mathrsfs}
\usepackage{latexsym}
\usepackage{showlabels}
\usepackage{color}
\usepackage{tensor}

\usepackage{accents}
\usepackage{nicefrac}
\newcommand{\tfrac}[2]{\nicefrac{#1}{#2}}

\newcommand{\pdv}[2]{\frac{\partial #1}{\partial #2}}
\newcommand{\JBessel}[3]{J_{#1}\left(\frac{#2}{\hbar}#3\right)}
\newcommand{\KBessel}[3]{K_{#1}\left(\frac{#2}{\hbar}#3\right)}
\newcommand{\braket}[2]{\left< #1 | #2 \right>}
\newcommand{\abs}[1]{\left| #1\right|}
\newcommand{\norm}[1]{\left| \left| #1 \right|\right|}
\newcommand{\ddd}[1]{\frac{\mathrm{d} #1}{2\pi}}
\usepackage{psfrag}
\newcommand{\eqref}[1]{(\ref{#1})}
\usepackage{graphicx}
\usepackage{subcaption}
\newcommand{\expval}[1]{\left< #1 \right>}
\usepackage[british]{babel}
\usepackage{afterpage}
\usepackage{hyperref} 
\newcommand{\sgn}[1]{\textrm{sgn}\left( #1 \right)}
\usepackage{mcite} 

\def\ben{\begin{equation}}
\def\een{\end{equation}}
\def\bena{\begin{eqnarray}}
\def\eena{\end{eqnarray}}

\def\bR{\mathbb{R}}

\def\dd{\mathrm{d}}

\def\H{\mathcal{H}}
\def\O{\mathcal{O}}
\begin{document}

\title{Singularity resolution depends on the clock}

\author{Steffen Gielen${^1}$, Luc\'ia Men\'endez-Pidal${^2}$}
\address{%
  ${^1}$School of Mathematics and Statistics,
  University of Sheffield,
  \\
  Hicks Building,
  Hounsfield Road,
  Sheffield S3 7RH,
  United Kingdom
  \\[.5em]
  ${^2}$School of Mathematical Sciences,
  University of Nottingham,
  \\
  University Park,
  Nottingham NG7 2RD,
  United Kingdom
}
\ead{s.c.gielen@sheffield.ac.uk, lucia.menendez-pidal@nottingham.ac.uk}

\begin{abstract}
We study the quantum cosmology of a flat Friedmann--Lema\^{i}tre--Robertson--Walker universe filled with a (free) massless scalar field and a perfect fluid that represents radiation or a cosmological constant whose value is not fixed by the action, as in unimodular gravity. We study two versions of the quantum theory: the first is based on a time coordinate conjugate to the radiation/dark energy matter component, i.e., conformal time (for radiation) or unimodular time. As shown by Gryb and Th\'ebault, this quantum theory achieves a type of singularity resolution; we illustrate this and other properties of this theory. The theory is then contrasted with a second type of quantisation in which the logarithm of the scale factor serves as time, which has been studied in the context of the ``perfect bounce'' for quantum cosmology. Unlike the first quantum theory, the second one contains semiclassical states that follow classical trajectories and evolve into the singularity without obstruction, thus showing no singularity resolution. We discuss how a complex scale factor best describes the semiclassical dynamics. This cosmological model serves as an illustration of the problem of time in quantum cosmology.
\end{abstract}


\maketitle

\section{Introduction}

One of the strongest motivations for considering  a quantisation of gravity arises from the spacetime singularities encountered in classical general relativity. Among such singularities, the perhaps most worrying one is the Big Bang of standard cosmology, which was presumably relevant for explaining the initial conditions of our Universe. The Big Bang is characterised by a vanishing spatial volume element, or determinant of the spatial metric, in the time slicing given by the constant curvature slices of a Friedmann--Lema\^itre--Robertson--Walker (FLRW) universe. One of the earliest questions for quantum gravity was whether the cosmological singularity can be avoided in a quantum theory of FLRW universes \cite{Misner1969}. Such {\em minisuperspace} models with finitely many degrees of freedom have long been used to study basic questions of quantised gravitational systems in the absence of a well-defined (continuum) realisation of the canonical quantisation programme for full general relativity \cite{DeWitt1967}. Our aim is to contribute to this long-standing debate by analysing a cosmological model with some peculiar and interesting features: in particular, we will show that the answer to whether the classical singularity is resolved in quantum cosmology depends on the choice of clock variable.

Classically, minisuperspace models are covariant under refoliations or redefinitions of the time variable, the only nontrivial diffeomorphisms left after reducing to FLRW symmetry. There is then no {\em a priori} preferred time parameter in the quantum theory -- leading to the {\em problem of time} known generally in canonical quantum gravity \cite{Isham1992,Kuchar1991,Anderson2012}. One of the multiple aspects of this problem, which has been studied in many examples in the literature \cite{Malkiewicz2019, *Malkiewicz2016}, is the inequivalence of quantum theories obtained by choosing different time coordinates before quantisation; this {\em multiple choice} problem  \cite{Kuchar1991} may be interpreted as a breaking of general covariance at the quantum level. To avoid this issue, one can try to work with a ``clock-neutral'' (in the terminology of \cite{Vanrietvelde2018, *Hoehn2018, *Hoehn2018_2}) Dirac quantisation, in which no time parameter is chosen before quantisation, quantum constraints are implemented on an initial (too large) kinematical Hilbert space and used to construct a physical inner product. Observables must then be compatible with the constraints, meaning they are ``frozen in time'': dynamics must be formulated relationally, in terms of observables corresponding to the value of quantity A when quantity B takes a given value \cite{Dittrich2006,*Dittrich2007,Rovelli1990,*Rovelli1991,Tambornino2011}. While conceptually appealing as a way of preserving general covariance, as with any other type of quantisation one has to check on a case-by-case basis how or whether Dirac quantisation can be implemented successfully. It is therefore insightful to study specific examples if possible.

The quantum cosmology model we study in this paper can be motivated from different directions. The first is to start from a flat FLRW universe filled with a (free) massless scalar field, one of the most widely studied models in all of quantum cosmology \cite{Blyth1975,Bojowald2005,*Ashtekar2006,*Ashtekar2007}. One way of expressing the problem of time is to say that since the Hamiltonian is constrained to vanish for reparametrisation-invariant systems, quantum states cannot evolve in the ``time'' given by the reading of an arbitrary coordinate label. What happens if one does not impose this constraint? At least for a particular choice of lapse function, one obtains the dynamics of unimodular gravity, where the now arbitrary value of the ``energy'' results in the appearance of a cosmological constant. One may think of the unimodular time parameter corresponding to this lapse choice as preferred, although this does not resolve the problem of time \cite{Unruh1989}. In other words, the extension of the original model in which there was only a massless scalar interacting with gravity comes about because  removing the Hamiltonian constraint adds an additional dynamical degree of freedom -- a dynamical cosmological constant and its conjugate variable, (unimodular) time. The new degree of freedom enters linearly into the Hamiltonian, which then admits a standard Schr\"odinger quantisation. The properties of this Schr\"odinger quantum theory were recently studied in a series of papers by Gryb and Th\'ebault \cite{Gryb2019, *Gryb2019_2, Gryb2019_1}, in particular the apparently generic singularity resolution displayed by this model. 

We will revisit these results and contrast them with a second viewpoint on the same dynamical system: by a redefinition of variables, the unimodular gravity model discussed above is mathematically {\em identical} to a minisuperspace model of a flat FLRW universe filled with a (free) massless scalar field and a second matter component now describing a perfect fluid with equation of state $p=\frac{1}{3}\rho$, i.e., radiation. This second model has been proposed as a quantum cosmology undergoing a ``perfect bounce'' \cite{Gielen2016, *Gielen2017}, a quantum transition through the classical singularity at zero volume rather than a conventional bounce in which the Universe stops collapsing at a finite size. The work of \cite{Gielen2016} mainly focused on the properties of solutions to the Wheeler--DeWitt equation, but also proposed a Klein--Gordon-like inner product which is conserved under evolution with respect to the scale factor $a$. This choice of scale factor time is not the same as working with a ``Schr\"odinger'' time coordinate, the variable conjugate to the perfect fluid momentum appearing linearly in the Hamiltonian -- which for radiation would be conformal time. Thus, the perfect bounce proposal suggests a different quantisation of the system studied by Gryb and Th\'ebault. An obvious question is whether this second type of quantum theory has similar singularity resolution properties. We will answer it in the negative: rather, as suggested by the ``perfect bounce'' idea, we will see that states can be evolved all the way back to the cosmological singularity without encountering anything special; one can, for instance, construct semiclassical states following the classical singular trajectories to arbitrary accuracy. 

While appearing perfectly consistent with the principles of quantum theory, the ``perfect bounce'' theory is somewhat unusual. It does not have a conventional unitary time evolution generated by a Hamiltonian; we will see that the best approximation to its dynamics at a semiclassical level may be given in terms of a complex ``effective'' Schr\"odinger time, as proposed in \cite{Bojowald2011}.  It does not appear to be equivalent to a Dirac quantisation in the usual sense, and there is no Hamiltonian that is required to be self-adjoint. The properties of this second quantum theory illustrate how a basic question such as singularity resolution in quantum cosmology can depend on the choice of clock. 

A simpler model closely related to ours, in which the only matter content was a perfect fluid modelling dust, had previously been studied by Gotay and Demaret \cite{Gotay1984}. Our results extend the results of \cite{Gotay1984}  and support the general conjecture given there (see also \cite{Gotay1996}) that the quantum dynamics of cosmological models are always singularity-free for ``slow'' clocks (which encounter a classical singularity in finite time) but lead to a singularity for ``fast'' clocks which do not have this property. Our model was also studied in \cite{Bojowald2016} using effective methods rather than wave functions in a Hilbert space, where the authors also found inequivalent results from different clock choices.

We introduce the classical cosmological model and its solutions in Section \ref{class-mod}. Section \ref{qtum-theory} discusses the two possible definitions of the quantum theory we are interested in, using two different variables as clocks and defining the inner product and notion of unitarity according to these choices. In Section \ref{sing-res-sect} we show that, while the Schr\"odinger clock leads to a quantum theory with generic singularity resolution, choosing a clock related to the scale factor leads to a quantum theory which reproduces classical behaviour, including classical singularities, with arbitrary accuracy. We conclude in Section \ref{conclusio}.

\section{The classical model}
\label{class-mod}

In this section we discuss the classical properties of the cosmological model we are studying. The model contains, in addition to a free massless scalar field, a second degree of freedom that can be interpreted either as dark energy or as a radiation fluid (or by extension, an arbitrary perfect fluid with fixed equation of state). All classical solutions encounter a Big Bang or Big Crunch singularity. The dark energy/radiation component is conjugate to a preferred (``Schr\"odinger'') time variable. More details on various properties of this model can be found in \cite{Gryb2019, *Gryb2019_2, Gielen2016}.

\subsection{Universe with a cosmological constant}

We consider a homogeneous and isotropic FLRW universe with a cosmological constant and no spatial curvature. The universe contains matter given by a massless scalar field $\phi$. The metric can be written as
\begin{equation}
\dd  s^2= -N(t)^2\dd t^2 + a(t)^2(\dd x^2+\dd y^2 +\dd z^2),
\label{metric}
\end{equation}
where $N(t)$ is the lapse function and we have set $c=1$. We start by considering the Einstein--Hilbert action together with the free action of a massless field:
\begin{equation}
\mathcal{S}=\mathcal{S}_{EH}+\mathcal{S}_\phi=\frac{1}{2\kappa}\int \dd^4x \ \sqrt{-g}\,[ R-2\Lambda] -\frac{1}{2}\int \dd^4x \ \sqrt{-g}\, g^{ab}\partial_a \phi \partial_b \phi
\end{equation}
where $\sqrt{-g}$ is the determinant of the spacetime metric $g_{ab}$, $R$ the Ricci scalar, and $\kappa=8\pi G$ where $G$ is the universal gravitational constant. The integral is in principle over the whole spacetime, but in order to avoid divergences it is convenient to restrict ourselves to an integration region $\Sigma\times \bR$, where $\Sigma$ is a compact space manifold and $\bR$ parametrises the time coordinate. For an infinite universe $\Sigma$ is usually viewed as a ``fiducial cell''.
After imposing isotropy and homogeneity \eref{metric} and $\phi=\phi(t)$, we find that the action simplifies (after discarding a total derivative) to
\begin{equation}
\mathcal{S}=\mathcal{S}_{EH}+\mathcal{S}_\phi=\int_\bR \dd t \left[ -\frac{3V_0 a}{\kappa N}\dot{a}^2-\frac{V_0a^3 N\Lambda}{\kappa}+\frac{V_0a^3}{2N}\dot{\phi}^2\right],
\label{action-cc}
\end{equation}
where $V_0=\int_\Sigma \dd^3 x <\infty$ is the coordinate volume of $\Sigma$ and the dot $\cdot$ represents $\frac{\dd}{\dd t}$. 

We now extend this theory so that it contains an additional dynamical degree of freedom: we promote $\Lambda$ from a parameter in the action to a dynamical variable $\Lambda(t)$. The perspective we would like to adopt is that $\Lambda$ is a constant of motion rather than a constant of nature, and as such can take different values for different classical solutions. One way to achieve this is to add an extra term to the action \eqref{action-cc},
\begin{equation}
\mathcal{S}=\mathcal{S}_{EH}+\mathcal{S}_\phi+\mathcal{S}_{cons}=\int_\bR \dd t \left[ -\frac{3V_0 a}{\kappa N}\dot{a}^2-\frac{V_0a^3 N\Lambda}{\kappa}+\frac{V_0a^3}{2N}\dot{\phi}^2+\Lambda\dot{T}\right]
\label{action-um}
\end{equation}
where we have introduced a Lagrange multiplier $T$ enforcing the constraint $\dot\Lambda=0$ (here and below we will assume variations defined such that Lagrange multipliers are held fixed at the spacetime boundary).

The gravitational action $\mathcal{S}_{EH}+\mathcal{S}_{cons}$ is equivalent to a parametrised version of unimodular gravity. Unimodular gravity is usually defined as a version of general relativity in which the metric determinant is fixed, $\sqrt{-g}=\epsilon$ where $\epsilon$ is a fixed volume form. This theory is then only invariant under the group of coordinate transformations that leave $\sqrt{-g}$ unchanged, which is smaller than the full diffeomorphism group of usual general relativity. Only the trace-free part of the Einstein equations is imposed as equations of motion \cite{Einstein1919,*Ellis2011}. In the Hamiltonian formulation, there is then no Hamiltonian constraint and $\Lambda$ appears as an integration constant corresponding to the value of the Hamiltonian on a given solution. 

A parametrised version of unimodular gravity is obtained by introducing additional fields in order to restore the full diffeomorphism symmetry into unimodular gravity. This leads to an action \cite{Henneaux1989,*Smolin2009}
\begin{equation}
\mathcal{S}_{PUM}=\frac{1}{2\kappa}\int \dd^4x \ \sqrt{-g}\,[ R-2\Lambda]+\Lambda\partial_\mu T^\mu
\end{equation}
where $\Lambda$ and $T^\mu$ are dynamical fields. One can obtain this action from an explicit ``parametrisation'' of unimodular gravity \cite{Kuchar1991b} in which a preferred set of coordinates satisfying the unimodular condition $\sqrt{-g}=\epsilon$ is promoted to dynamical fields $X^A(x)$ where the $x$ are now arbitrary coordinate labels. It is clear that the restriction of the additional term $\Lambda\partial_\mu T^\mu$ to FLRW symmetry leads to the term $\Lambda\dot{T}$ added in \eqref{action-um} and that we are hence working with a parametrised form of unimodular gravity.

While this parametrised form has full diffeomorphism symmetry and allows for arbitrary choices of lapse $N$, below we will work in a unimodular gauge for $N$ in which the dynamics take particularly simple form, the one of usual unimodular gravity. In this gauge the lapse satisfies $N(t)\propto a(t)^{-3}$ (see, e.g., \cite{Nojiri2016}).

In the Hamiltonian theory for \eqref{action-um}, the conjugate momenta to $a$ and $\phi$ are then
\begin{equation}
\pi_a = -\frac{6V_0a}{\kappa N}\dot{a}, \hspace{5mm}
\pi_\phi = \frac{V_0a^3}{N}\dot{\phi},
\label{momenta}
\end{equation}
while $T$ and $\Lambda$ are conjugate, $\{T,\Lambda\}=1$; the Hamiltonian is
\begin{equation}
\H=N\left[-\frac{\kappa}{12V_0a}\pi_a^2+\frac{1}{2V_0 a^3}\pi_\phi^2 + \frac{V_0 a^3}{\kappa}\Lambda\right].
\label{ham-cc}
\end{equation}
In the following we set $\kappa=8\pi G=1$ to simplify the notation.  The lapse $N$ is an undetermined function of time and can be regarded as a Lagrange multiplier. This imposes the additional constraint, usually called Hamiltonian constraint
\begin{equation}
C=-\frac{1}{12V_0a}\pi_a^2+\frac{1}{2V_0 a^3}\pi_\phi^2 + V_0 a^3\Lambda=0.
\label{constraint-cc}
\end{equation}

At this point, the addition of a conservation constraint $\Lambda\dot{T}$ may appear somewhat {\em ad hoc}. However we will see shortly that our construction here can also be seen as adding a perfect fluid as additional matter, parametrised by the variables $T$ and $\Lambda$. We would then choose an equation of state $p=-\rho$ for this perfect fluid, which corresponds to dark energy. We will see below that other equations of state are equally possible.

With the change of variables
\begin{equation}
v=\frac{2}{\sqrt{3}}V_0^{\tfrac{1}{2}}a^3, \hspace{4mm} \varphi=\sqrt{\frac{3}{2}}\phi, \hspace{4mm} \pi_v=V_0^{-\tfrac{1}{2}}\sqrt{\frac{1}{12}}\frac{\pi_a}{a^2}, \hspace{4mm} \pi_\varphi=\sqrt{\frac{2}{3}}\pi_\phi
\label{v-varphi}
\end{equation}
the Hamiltonian \eqref{ham-cc} simplifies to
\begin{equation}
\H=M\left[-\pi_v^2 +\frac{\pi_\varphi^2}{v^2}+\lambda\right],
\label{ham-cc-sim}
\end{equation}
where $\lambda=V_0\Lambda$ and $M=\frac{\sqrt{3}}{2}N  v V_0^{-\tfrac{1}{2}}$ are the rescaled cosmological constant and redefined lapse function. The variable $v$ is proportional to the three-volume of $\Sigma$. Working in the unit convention where the line element has units of length and the metric is dimensionless, we see that $[\H]=L$ (writing everything in terms of powers of length) and $M$ is dimensionless. In fact, the only dimensionful constant left is $V_0$ with $[V_0]=L^3$. In these variables the Hamiltonian constraint becomes
\begin{equation}
\mathcal{C}=-\pi_v^2 +\frac{\pi_\varphi^2}{v^2}+\lambda=0.
\label{ham-constraint-cc-simp}
\end{equation}

Defining a metric $\eta_{AB}=\textrm{diag}(-1,v^2)$, the resulting Hamiltonian is of the form 
\begin{equation}
\mathcal{H}=M[\eta^{AB}p_Ap_B+\lambda],
\label{ham-metric}
\end{equation}
where $p_A$ represents the momenta $\pi_v$ and $\pi_\varphi$. $\eta_{AB}$ is the metric of a portion of two dimensional flat space covered by the coordinate chart $(v,\varphi)$. If one views $v$ as spacelike and $\varphi$ as timelike, one would refer to this part of Minkowski spacetime as the Rindler wedge; if $v$ is defined as timelike and $\varphi$ as spacelike one obtains the future light cone of a point in Minkowski space, which one might call the Milne wedge (or $(1+1)$-dimensional Milne universe). At this point either terminology would be equally appropriate, but since later in the paper we will argue for the volume $v$ as a time coordinate it might be more useful to think of the Milne wedge. The boundary of the Milne wedge is at $v=0$; the Milne wedge is not geodesically complete as geodesics can reach the Big Bang/Big Crunch $v=0$ in finite time. Another important property of \eqref{ham-metric} is that the Hamiltonian is quadratic in all momenta except $\lambda$ for which it is linear.

\begin{figure}[ht]
\centering
\begin{subfigure}{0.32\textwidth}
\centering
\includegraphics[scale=0.4]{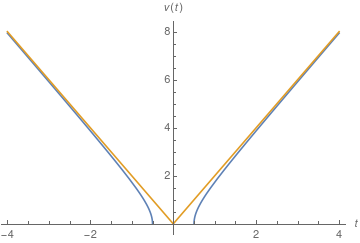}
\caption{$\lambda =1$}
\end{subfigure}
\begin{subfigure}{0.32\textwidth}
\centering
\includegraphics[scale=0.4]{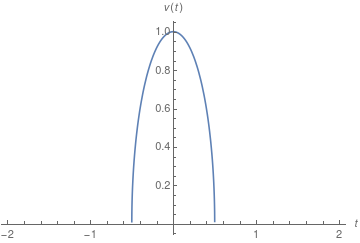}
\caption{$\lambda=-1$}
\end{subfigure}
\begin{subfigure}{0.32\textwidth}
\centering
\includegraphics[scale=0.4]{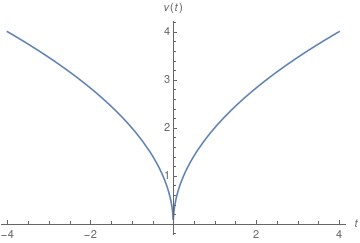}
\caption{$\lambda=0$}
\end{subfigure}
\caption{The classical solution $v(t)$ (for $t_0=\varphi_0=0$ and $\pi_\varphi=1$) depending on the sign of the cosmological constant. The yellow lines in $(a)$ represent the asymptotes $v(t)= 2\sqrt{\lambda}|t|$. This universe expands or contracts forever. In the negative $\lambda$ case $v(t)$ reaches a maximum and then has a turnaround.
The case $(c)$ expands or contracts forever too, but with different asymptotic behaviour compared to $(a)$.}
\label{classical-trajectories}
\end{figure}

\begin{figure}[ht]
\centering
\begin{subfigure}{0.32\textwidth}
\centering
\includegraphics[scale=0.4]{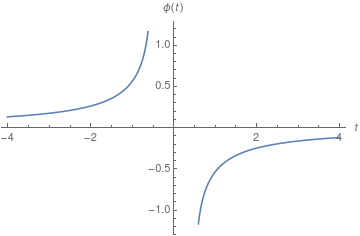}
\caption{$\lambda =1$}
\end{subfigure}
\begin{subfigure}{0.32\textwidth}
\centering
\includegraphics[scale=0.4]{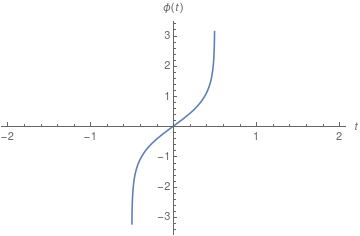}
\caption{$\lambda=-1$}
\end{subfigure}
\begin{subfigure}{0.32\textwidth}
\centering
\includegraphics[scale=0.4]{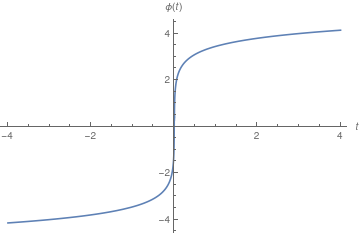}
\caption{$\lambda=0$}
\end{subfigure}
\caption{The classical solution $\varphi(t)$ (for $t_0=\varphi_0=0$ and $\pi_\varphi=1$) depending on the sign of the cosmological constant. The field $\varphi$ diverges at the singular points of the metric. Note that $t_0=0$ is not well defined for $\varphi(t)$ in \eqref{class-sol-v-varphi-1}, but can be approximated. In our case we used the value $t_0=10^{-3}$ to generate the plot.}
\label{classical-trajectories-2}
\end{figure}

A particularly simple form of the dynamics is obtained for a time coordinate defined by $M=1$. In terms of the lapse $N$, we then have $N=\frac{2}{\sqrt{3}}\frac{V_0^{\tfrac{1}{2}}}{ v}$ which is a unimodular time: the metric has determinant $-1$. This coordinate choice reduces the dynamics to the usual form of unimodular gravity. Applying the Hamiltonian formalism to the system and assuming $\pi_\varphi\neq 0$, we find the  classical solutions
\begin{equation}
v(t)=\sqrt{-\frac{\pi_\varphi^2}{\lambda}+4\lambda(t-t_0)^2}, \hspace{3mm} \varphi(t)=\frac{1}{2}\log\left|\frac{\pi_\varphi-2\lambda(t-t_0)}{\pi_\varphi+2\lambda(t-t_0)}\right|+ \varphi_0
\label{class-sol-v-varphi}
\end{equation}
for $\lambda\neq 0$. The parameters $t_0$ and $\varphi_0$ are constants of integration, and $\pi_\varphi$ is also a constant of motion. We see that $v=0$ for $t-t_0=\pm\frac{\pi_\varphi}{2\lambda}$. If $\lambda=0$ the solutions are
\begin{equation}
v(t)=2\sqrt{|\pi_\varphi| |t-t_0|}, \hspace{5mm} \varphi(t)=\sgn{\pi_\varphi(t-t_0)}\frac{\log|\frac{t}{t_0}-1|}{2}+\varphi_0.
\label{class-sol-v-varphi-1}
\end{equation}
The solutions are plotted in Figures \ref{classical-trajectories} and \ref{classical-trajectories-2} for different signs of $\lambda$.

Looking at \eqref{ham-metric} with $M=1$, we see that any variable $\tilde{t}$ conjugate to $\lambda$ satisfies $\frac{\dd \tilde{t}}{\dd t}=1$, i.e., $\tilde{t}=t$ up to a constant (see also \cite{Henneaux1989, *Smolin2009}). In this sense, the cosmological constant $\lambda$ is conjugate to a unimodular time coordinate. This will allow us to build relational observables which correspond in a close sense to the classical solutions \eqref{class-sol-v-varphi} and \eqref{class-sol-v-varphi-1}.

This universe contains an initial singularity, for all values of $\lambda$ and all $\pi_\varphi\neq 0$. A quick way to acknowledge this is to focus on comoving observers ($\dd x = \dd y = \dd z =0$). Curvature invariants diverge as $v\rightarrow 0$, and so if $v=0$ can be reached in finite proper time by comoving observers it is safe to say that the spacetime has a singularity. Let us consider first $\lambda>0$ and the expanding branch $t>0$. Then, setting $t_0=0$ for simplicity, with $t_{sing}=\frac{|\pi_\varphi|}{2\lambda}$ and $t_1>t_{sing}$ the proper time for comoving observers is
\begin{equation}
\int_{\tau_{sing}}^{\tau_1} \dd \tau=2\sqrt{\frac{V_0}{3}}\int_{t_{sing}}^{t_1} \frac{\dd t}{v(t)}=\sqrt{\frac{V_0}{3\lambda}}\log\left(\frac{2\lambda t_1}{|\pi_\varphi|}+\sqrt{\frac{4\lambda^2t_1^2}{\pi_\varphi^2}-1}\right)< \infty.
\end{equation}
Comoving observers reach $v=0$ in a finite proper time. This calculation can be appropriately adapted to the $\lambda<0$ case. If $\lambda=0$ we have
\begin{equation}
\int_{\tau_{sing}}^{\tau_1} \dd \tau=2\sqrt{\frac{V_0}{3}}\int_{t_{sing}}^{t_1} \frac{\dd t}{v(t)}= 2\sqrt{\frac{V_0}{3|\pi_\varphi|}}\sqrt{t_1}<\infty.
\end{equation}
This initial singularity is referred to as Big Bang if it is in the past of the comoving observer and Big Crunch if it is in the future. Due to the symmetry of the model this universe contains both.

The unimodular time parameter $t$ is a good choice of clock; it parametrises entire solutions from the singularity out to infinity. As we explained, this time variable is essentially the same as the dynamical variable conjugate to $\lambda$ -- which in the following we also refer to as $t$. All other variables can be expressed as a function of $t$ globally. From \eqref{class-sol-v-varphi} and \eqref{class-sol-v-varphi-1} we can also infer that in the case $\lambda\geq 0$, $v$ (or equivalently $\log(v/v_0)$) and $\varphi$ are good clocks. Indeed, in the $t$ range for which these variables are well-defined, both of them are monotonic functions on each branch and therefore can parametrise the rest of the variables injectively (see figures \ref{classical-trajectories} and \ref{classical-trajectories-2}). More attention should be paid in the $\lambda<0$ case, where the volume $v$ reaches a local maximum and turns around. In the neighbourhood of this maximum $v$ is not a good clock any more. In classical physics all clocks are equivalent in the sense that, at least in regions where these are all monotonic functions, we may equally well express a classical solution in terms of $v(t)$, $v(\varphi)$ or $t(v)$. There is an underlying unique theory which is coordinate-independent. In general this ``clock invariance'' is not conserved in the quantum theory, as we will see in this paper.

\subsection{Universe with radiation}

In the previous subsection we discussed a model in which the cosmological constant appears as a constant of motion rather than a constant of nature, which one can interpret as defining gravity by unimodular gravity instead of general relativity. Here we will interpret the same equations of motion as describing a different universe in which there is no cosmological constant but the additional term we added gives a matter contribution to the action.

We consider the same FLRW metric \eqref{metric}, without a cosmological constant but with a radiation perfect fluid and the same massless scalar field $\phi$. The radiation fluid is parametrised by the energy density $\rho=ma^{-4}$ where $m$ is a constant of motion. (Note that $m$ and $\rho$ both have units of energy density.) This model might be closer to a realistic cosmology than the previous one, since the early universe was dominated by radiation whereas dark energy was negligible. Indeed this was one of the main motivations for its study in the ``perfect bounce'' cosmology in \cite{Gielen2016}. For this model the action is
\begin{equation}
\mathcal{S}=\mathcal{S}_{EH}+\mathcal{S}_\phi+\mathcal{S}_{rad}=\int_\bR \dd t \left[-\frac{3V_0a}{\kappa N}\dot{a}^2+ \frac{V_0a^3}{2N}\dot{\phi}^2-\frac{V_0N}{a}m+m\dot{\chi}\right].
\label{action-with-radiation}
\end{equation}
The contribution of the radiation fluid can be obtained by starting with the general action for an isentropic fluid \cite{Brown1993} $\mathcal{S}_{isen}=\int \dd^4 x [-\sqrt{-g}\, \rho\left(\frac{|J|}{\sqrt{-g}}\right)+ J^a\partial_a \varphi' +\beta_A \partial_a\alpha^AJ^a]$ where $\varphi'$, $\beta_A$ and $\alpha^A$ are spacetime scalars and $J^a$ is the densitised particle number flux vector (and the energy density $\rho$ is initially a function of $J^a$ and $\sqrt{-g}$). The last two terms in $\mathcal{S}_{isen}$ generate  constraints on the flux vector $J^a$ after varying with respect to $\varphi'$, $\beta_A$ and $\alpha^A$. 
After restriction to FLRW symmetry, the term $J^a\partial_a \varphi'$ can be written as $m\dot\chi$ for an appropriately redefined Lagrange multiplier $\chi$, while the second term can be dropped since in FLRW symmetry the fluid flow is automatically orthogonal to constant time hypersurfaces, and no further constraints on it are needed. Thus we obtain \eqref{action-with-radiation}. We now also see that the addition of a term $m\dot{\chi}$ enforcing energy conservation on $m$ is similar to what we did in \eqref{action-um} for the cosmological constant.

The Hamiltonian resulting from \eqref{action-with-radiation} is now
\begin{equation}
\H=N\left[-\frac{\kappa }{12V_0a}\pi_a^2+\frac{1}{2V_0a^3}\pi_\phi^2 + \frac{V_0}{a}m\right].
\label{ham-re}
\end{equation}
As in the previous subsection, the Hamiltonian is constrained to vanish, and this constraint arises from variation with respect to the lapse $N$. The variable $m$ now plays a very similar r\^ole to $\Lambda$ previously. In terms of physical motivation, one might previously have considered the cosmological constant $\Lambda$ to take either sign but choosing $m<0$ might not be as well motivated, as it would correspond to radiation with negative energy density. After setting again $\kappa=1$, we perform the change of variables (here just a simple rescaling)
\begin{equation}
\alpha=2\sqrt{3}V_0^{\tfrac{1}{2}}a, \hspace{5mm} \vartheta=\sqrt{\frac{1}{6}}\phi, \hspace{5mm} \pi_\alpha=\sqrt{\frac{1}{12}}V_0^{-\tfrac{1}{2}}\pi_a, \hspace{5mm} \pi_\vartheta =\sqrt{6}\pi_\phi,
\label{varchange2}
\end{equation}
and find that the Hamiltonian has the form
\begin{equation}
\H=\tilde{N}\left[-\pi_\alpha^2+\frac{\pi_\vartheta^2}{\alpha^2}+\tilde{m}\right]
\label{ham-re-sim}
\end{equation}
where $\tilde{N}=\frac{2\sqrt{3} N V_0^{\tfrac{1}{2}}}{\alpha}$ and $\tilde{m}=V_0m$ (which has units of mass). Again, $\tilde{m}$ is a dynamical variable (and constant of motion) of the theory, conjugate to another variable $\eta=\frac{\chi}{V_0}$. 

Clearly, when written in these variables the Hamiltonian is mathematically identical to \eqref{ham-cc-sim} and therefore every result regarding one model can be transferred to the other one ($\alpha$ should be positive definite just as $v$ was previously). The preferred lapse choice $\tilde{N}=1$ now corresponds to $N\propto a$, i.e., working in conformal time: classical solutions for a universe with dark energy written in unimodular time correspond to solutions with radiation in conformal time. In the following we will be working with the variables $v$, $\varphi$ and $\lambda$ but it will be understood that all results can be translated to the perfect bounce by identifying them respectively with $\alpha$, $\vartheta$ and $\tilde{m}$.  We can again view \eqref{ham-re-sim} as defining a Hamiltonian for a particle moving in the Milne wedge, cf.~\eqref{ham-metric}.

We are primarily interested in a matter component that can be interpreted either as dark energy or as radiation, but the above discussion can be extended to any perfect fluid with equation of state $p=w\rho$ with $w< 1$. In this more general case, the energy density has the form $\rho = m a^{-3(1+w)}$ where $m$ is conserved and the Hamiltonian is
\begin{equation}
\H=N\left[-\frac{\kappa }{12V_0a}\pi_a^2+\frac{1}{2V_0a^3}\pi_\phi^2 + \frac{V_0}{a^{3w}}m\right].
\end{equation}
This can now be brought into the canonical form \eqref{ham-re-sim} by the variable change ($\kappa=1$)
\begin{eqnarray}
\alpha&=&\frac{4}{\sqrt{3}}V_0^{\tfrac{1}{2}}\frac{a^{\frac{3(1-w)}{2}}}{1-w},\quad\pi_\alpha=\sqrt{\frac{1}{12}}V_0^{-\tfrac{1}{2}}\pi_a\,a^{\frac{3w-1}{2}},\nonumber\\
\vartheta &=& \sqrt{\frac{3}{8}}(1-w)\phi,\quad \pi_\vartheta = \frac{\sqrt{8}}{\sqrt{3}(1-w)}\pi_\phi
\end{eqnarray}
and, as before, $\tilde{m}=V_0m$. The preferred lapse choice $\tilde{N}=1$ now corresponds to $N\propto a^{3w}$. The expressions we gave for dark energy ($w=-1$) in \eqref{v-varphi} and for radiation ($w=\frac{1}{3}$) in \eqref{varchange2} are special cases of this more general construction. The new canonical variable $\alpha$ is, in the general case, a fractional power of the scale factor whose interpretation is less obvious than for a volume factor or scale factor in the cases we consider in the paper. However, since both $a$ and $\alpha$ are required to be nonnegative, the mathematical equivalence of the two theories we have discussed so far clearly extends to the more general case of a perfect fluid with arbitrary equation of state. In particular, quantum cosmological dust models with $w=0$ such as have been studied in \cite{Gotay1984, Ali2018, *Husain2019} fit naturally into our formalism, and would be subject to the same quantisation ambiguities we discuss here. For dust, the preferred time coordinate is proportional to proper time \cite{Brown1994} and the canonical variable for the geometry would be proportional to $a^{3/2}$.

\subsection{Dirac observables}

Our dynamical system has a gauge symmetry under time reparametrisations, which implies that not all phase space variables correspond to physical observables. Defining a sufficiently large set of observables is however important to characterise and distinguish different trajectories of this system. Here we focus on the notion of Dirac observables commonly used in the Hamiltonian description of constrained systems \cite{diracbook}.

The Hamiltonian constraint \eqref{ham-constraint-cc-simp} defines a constraint surface $\mathscr{C}$ in the phase space of the model where the solutions to the equation of motion evolve. Dirac observables are functions $O$ on the constraint surface that are invariant under the flow generated by the Hamiltonian constraint $\mathcal{C}$:
\begin{equation}
\lbrace  \mathcal{C},O \rbrace \approx 0,
\end{equation}
where $\lbrace \cdot, \cdot \rbrace$ is the Poisson bracket defined by the momenta and coordinates of the phase space and the $\approx$ sign is Dirac's notion for an equality that only needs to hold in $\mathscr{C}$. Concretely, for our model we have
\begin{equation}
\lbrace \mathcal{C},O \rbrace= -\frac{2\pi_\varphi^2}{v^3}\pdv{O}{\pi_v}-\pdv{O}{t}-\frac{2\pi_\varphi}{v^2}\pdv{O}{\varphi}+2\pi_v\pdv{O}{v},
\end{equation}
and it is already obvious that $\pi_\varphi$ and $\lambda$ (which are constants of motion) and any combination of the two are Dirac observables. 

The time reparametrisation invariance of our cosmological model is a remnant of full diffeomorphism symmetry. In particular, the dynamical Hamiltonian is itself a multiple of the constraint $\mathcal{C}$. This is why any Dirac observable in our system is also a constant of motion. This observation has been made from a slightly different perspective, e.g., in \cite{Unruh1989}, and can be seen as one facet of the problem of time: it appears that the only observable quantities are ones that are constant under the entire cosmological evolution.

It is then convenient to introduce an important family of Dirac observables, namely relational observables \cite{Dittrich2006,*Dittrich2007,Rovelli1990,*Rovelli1991,Tambornino2011}. These quantities give information on how two phase space functions change with respect to one another. Two examples of this type of observables are $v(t=t_1)$ and $t(v=v_1)$, or the volume element when $t$ takes the value $t_1$ and the time $t$ when the volume element takes the value $v_1$. By varying the values of $t_1$ and $v_1$ on the allowed range we can build a family of observables; each of these will be a constant along each dynamical solution, but together they define {\em complete observables} which contain information about the full dynamics of the system.

To define these observables, start from an initial point $P$ on the constraint surface $\mathscr{C}$ whose coordinates are $(t_{{\rm i}}, v_{{\rm i}}, \varphi_{{\rm i}}, \lambda, \pi_{v_{{\rm i}}}, \pi_\varphi )$. For $P$ to lie in $\mathscr{C}$ we must have $-\pi_{v_{{\rm i}}}^2+\frac{\pi_\varphi^2}{v_{{\rm i}}^2}+\lambda=0$. We label the initial values with subscript ${}_{{\rm i}}$ apart from $\lambda$ and $\pi_\varphi$, the constants of motion. The initial data determine a unique classical solution within $\mathscr{C}$; one can then ask for this solution what the value of the volume $v$ is at the ``time'' for which a $t$ clock shows the value $t_1$. The explicit expressions for these observables are 
\begin{equation}
v(t=t_1)=\left\lbrace \begin{array}{cc}
\sqrt{-\frac{\pi_\varphi^2}{\lambda}+4\lambda\left(t_1-t_{{\rm i}}-\frac{v_{{\rm i}}\pi_{v_{{\rm i}}}}{2\lambda}\right)^2}, & \lambda<0 \ \textrm{or} \ \lambda >0\\[3mm]
 2\sqrt{|\pi_\varphi|\left|t_1-t_{{\rm i}}+\frac{v_{{\rm i}}^2}{4\pi_\varphi}\right|},& \lambda=0.
\end{array} \right.
\label{voft}
\end{equation} 
The first expression is only valid when the argument of the square root is positive; as we saw earlier, the classical solutions reach singularities at some finite value of $t$ beyond which they would not be defined. Inverting the expressions and choosing only the expanding branch of the solution, we find
\begin{equation}
t(v=v_1)=\left\lbrace \begin{array}{cc}
t_{{\rm i}}+\frac{v_{{\rm i}}\pi_{v_{{\rm i}}}}{2\lambda}+\frac{1}{2}\sqrt{\frac{v_1^2}{\lambda}+\frac{\pi_\varphi^2}{\lambda^2}}, & \lambda<0 \ \textrm{or} \ \lambda >0\\[3mm]
\frac{v_1^2-v_{{\rm i}}^2}{4|\pi_\varphi|}+t_{{\rm i}},& \lambda=0 .
\end{array} \right. 
\label{tofv}
\end{equation} 
Again, the plots in Fig.~\ref{classical-trajectories} show that there would be two different values for $t$ corresponding to the same value for $v$, so we need to make a choice.

If $\lambda<0$, the first expression in (\ref{tofv}) is only valid when $v_1\le\frac{|\pi_\varphi|}{\sqrt{-\lambda}}$, which is the upper bound reached by the volume in this case.  The expressions given for $v(t=t_1)$ and $t(v=v_1)$, when seen as functions of their argument, are special cases of the general solutions given in (\ref{class-sol-v-varphi})-(\ref{class-sol-v-varphi-1}), showing that these remain in the constraint surface $\mathscr{C}$ for all values of the argument. The relational observables $v(t=t_1)$ and $t(v=v_1)$ only satisfy, e.g., $\lbrace \mathcal{C},v(t=t_1)\rbrace = f\,\mathcal{C}$ for some function $f\neq 0$ on phase space, and therefore $\lbrace \mathcal{C},v(t=t_1)\rbrace=0$ only in $\mathscr{C}$ and not everywhere in phase space, unlike for the simple constants of motion $\pi_\varphi$ and $\lambda$. The reason why the trajectories $v(t)$ and $t(v)$ can be used to define observables is essentially that, because of the appearance of a dark energy or radiation component, $t$ is a phase space coordinate rather than an arbitrary parameter.

\section{The quantum theory}
\label{qtum-theory}

In this section we study the quantisation of the model and compare the properties of Hilbert spaces constructed from several possible clock choices. Quantisation of the Hamiltonian (\ref{ham-metric}) leads to the Wheeler--DeWitt equation, which can be solved by a separation of variables ansatz and has Bessel functions as solutions. To give physical meaning to those solutions, we then choose one dynamical variable as a clock and construct a Hilbert space with positive definite inner product for the remaining variables. The choice of clock leads to different boundary conditions for the solutions and hence different quantum theories. For each clock choice singularity resolution is analysed.

In general, the definition of a Wheeler--DeWitt equation is subject to ordering ambiguities and ambiguities arising from a choice of coordinates on phase space. Here we follow the operator ordering prescription proposed by Hawking and Page \cite{Hawking1985}: one replaces the term $\eta^{AB}p_Ap_B$ of \eqref{ham-metric} which is quadratic in momenta by $-\hbar^2\square$ where $\square$ is the Laplace--Beltrami operator
\begin{equation}
\square= \frac{1}{\sqrt{-\eta}}\partial_A(\eta^{AB}\sqrt{-\eta}\partial_B)
\end{equation}
associated to the Minkowski metric $\eta_{AB}$.
This operator ordering ensures covariance of the theory under changing coordinates from $(v,\varphi)$ to different coordinates on the Milne wedge \footnote{General covariance does not fix this form entirely in general; for instance one could add a contribution $\hbar^2\xi R$ where $R$ is the Ricci scalar \cite{DeWitt1957, *Halliwell1988, *Moss1988}. In our case, since $\eta_{AB}$ is flat, such ambiguities do not arise.}; however, we will see shortly that self-adjointness of the $\square$ operator on the Milne wedge is not guaranteed and needs to be considered carefully. The Wheeler--DeWitt equation is then obtained by considering a wave function $\Psi(v,\varphi, t)$ and applying to it the operator resulting from the Hamiltonian with lapse choice $M=1$:
\begin{equation}
\left( \hbar^2\frac{\partial^2}{\partial v^2}+\frac{\hbar^2}{v}\pdv{}{v}-\frac{\hbar^2}{v^2}\frac{\partial^2}{\partial\varphi^2}-i\hbar \pdv{}{t}\right)\Psi(v,\varphi,t)=0.
\label{wdw}
\end{equation}

This equation can be solved  using a separation of variables ansatz of the form $\Psi(v,\varphi,t)=e^{i\lambda \frac{t}{\hbar}}\psi_\lambda(v)\nu(\varphi)$. This results in two equations,
\begin{equation}
\frac{\nu{''}(\varphi)}{\nu(\varphi)}=-k^2
\label{vawe-eq}
\end{equation}
for the $\varphi$ part, and
\begin{equation}
v^2\psi_\lambda''(v)+v\psi_\lambda'(v)+\left(\frac{\lambda}{\hbar^2}v^2 +k^2\right)\psi_\lambda(v)=0
\label{bessel-equation}
\end{equation}
for the $v$ part. Here the notation $'$ refers to the derivative of the function. \eqref{vawe-eq} has solutions of the form $\nu(\varphi)=e^{ik\varphi}$, where we restrict $k$ to be real since real exponentials will not be normalisable in the inner products of interest below. \eqref{bessel-equation} is known as Bessel's equation and has Bessel functions $J_{\pm ik}\left(\frac{\sqrt{\lambda}}{\hbar}v\right)$ as solutions. For concreteness, here and in the following we define these for negative $\lambda$ by
\begin{equation}
J_{ik}\left(\frac{\sqrt{\lambda}}{\hbar}v\right)=J_{ik}\left(i\frac{\sqrt{-\lambda}}{\hbar}v\right)=e^{-\frac{k\pi}{2}}I_{ik}\left(\frac{\sqrt{-\lambda}}{\hbar}v\right)
\label{besseldef}
\end{equation}
where $I_\alpha(z)$ is a modified Bessel function of the first kind. A general solution to the Wheeler--DeWitt equation is then of the form
\begin{equation}
\fl
\Psi(v,\varphi,t)=\int_{-\infty}^\infty \frac{\dd k}{2\pi} \int_{-\infty}^\infty \frac{\dd \lambda}{2\pi\hbar} \ e^{i\frac{\lambda}{\hbar}t}e^{ik\varphi}\left( \alpha(k,\lambda)\JBessel{ik}{\sqrt{\lambda}}{v}+\beta(k,\lambda)\JBessel{-ik}{\sqrt{\lambda}}{v}\right),
\label{gen-sol-wdw-1}
\end{equation}
which can alternatively be written as
\begin{eqnarray}
\fl
\Psi(v,\varphi,t)&=&\int_{-\infty}^\infty \frac{\dd k}{2\pi} \int_0^\infty \frac{\dd \lambda}{2\pi\hbar} \ e^{i\frac{\lambda}{\hbar}t}e^{ik\varphi}\left( \alpha(k,\lambda)\JBessel{ik}{\sqrt{\lambda}}{v}+\beta(k,\lambda)\JBessel{-ik}{\sqrt{\lambda}}{v}\right)
\label{gen-sol-wdw-2}
\\\fl
&&+\int_{-\infty}^\infty \frac{\dd k}{2\pi} \int_{-\infty}^0 \frac{\dd \lambda}{2\pi\hbar} \ e^{i\frac{\lambda}{\hbar}t}e^{ik\varphi}\left( \alpha(k,\lambda)\KBessel{ik}{\sqrt{-\lambda}}{v}+ \beta(k,\lambda)I_{ik}\left(\frac{\sqrt{-\lambda}}{\hbar}\right)\right).\nonumber
\end{eqnarray}
In (\ref{gen-sol-wdw-2}) we have rewritten the Bessel functions with imaginary argument in terms of the more commonly used modified Bessel functions $K$ and $I$, and redefined the functions $\alpha(k,\lambda)$ and $\beta(k,\lambda)$ for $\lambda<0$. We see explicitly that universes in a superposition of cosmological constant $\lambda$ are in principle possible. This is fundamentally different from treatments that do not come from unimodular gravity in which the cosmological constant is a constant of nature and not a dynamical variable. In the radiation density interpretation, this means that we can have superpositions of the mass parameter $m$ (which can take either positive or negative values at this stage).

To analyse these solutions and give physical meaning to them we need to construct a Hilbert space. In the deparametrised framework we are using here, this requires a choice of one of the dynamical variables to be the relational time parameter or clock of the system. As explained in Section \ref{class-mod}, either $t$, $v$ or $\varphi$  might be possible clock variables. Choosing a clock amounts to defining a quantum theory for the two remaining degrees of freedom as evolving in a parameter specifying the possible ``outcomes'' for the clock; the clock is then not associated with an operator on the corresponding Hilbert space.

\subsection{The Schr\"odinger clock $t$}
\label{t-clock}

Defining the Hamiltonian
\begin{equation}
\hat{\H}_s= \hbar^2\left( -\frac{\partial^2}{\partial v^2}-\frac{1}{v}\pdv{}{v}+\frac{1}{v^2}\frac{\partial^2}{\partial\varphi^2}\right),
\label{ham-qtum}
\end{equation}
the Wheeler--DeWitt equation \eqref{wdw} looks like a Schr\"odinger equation using $t$ as a time parameter:
\begin{equation}
i\hbar\pdv{}{t}\Psi(v,\varphi,t)=-\hat{\H}_s\Psi(v,\varphi,t).
\label{schrod}
\end{equation}
The possible values of $\lambda$ used in the plane wave decomposition (\ref{gen-sol-wdw-1}) now correspond to energy eigenvalues of $\hat{\H}_s$. This suggests defining a Schr\"odinger inner product
\begin{equation}
\braket{\Psi}{\Phi}=\int_0^\infty \dd v \int_{-\infty}^{\infty} \dd \varphi \; v\,\bar{\Psi}(v,\varphi,t)\Phi(v,\varphi,t)
\label{inner-prod-t}
\end{equation}
in the Hilbert space $L^2(\mathcal{M},\dd v \dd \varphi \sqrt{-\eta})$, where $\mathcal{M}$ is the Milne wedge and $\eta$ its associated metric. $\braket{\Psi}{\Phi}$ is positive definite and sesquilinear by construction and hence is a well-defined inner product. The preferred time variable $t$ (unimodular time if we use the dark energy interpretation, conformal time if we use the radiation interpretation of the model) now plays the r\^{o}le of a time parameter in non-relativistic quantum mechanics.

For this quantum theory to be consistent we would like to have unitary evolution: inner products $\braket{\Psi}{\Phi}$ should be conserved in time $t$. One can impose self-adjointness on $\hat{\H}_s$ or alternatively verify that
\begin{equation}
\pdv{}{t}\braket{\Psi}{\Phi}=0
\label{inner-prod-cons}
\end{equation}
for any two solutions to (\ref{schrod}). This condition leads to the boundary condition
\begin{equation}
\int \dd \varphi \left[ v\left( \bar{\Psi}\pdv{}{v}\Phi-\Phi\pdv{}{v}\bar{\Psi}\right)\right]_{v=0}^{v=\infty}=0.
\label{bound}
\end{equation}
Not all square integrable solutions to \eqref{wdw} satisfy \eqref{bound}; one needs to choose a subspace of $L^2(\mathcal{M},\dd v \dd \varphi \sqrt{-\eta})$ on which \eqref{bound} holds, and where the Hamiltonian $\hat{\H}_s$ can have a self-adjoint extension. More details on this procedure are given, e.g., in \cite{Gryb2019, *Gryb2019_2}. The appearance of boundary conditions for the Hamiltonian $\hat{\H}_s$ is well-known in standard quantum mechanics: after defining $\xi_\lambda=\sqrt{v}\psi_\lambda$ Bessel's equation \eqref{bessel-equation} becomes
\begin{equation}
-\hbar^2\frac{\partial^2}{\partial v^2}\xi_\lambda-\hbar^2\frac{\frac{1}{4}+k^2}{v^2}\xi_\lambda=\lambda\xi_\lambda,
\end{equation}
a Schr\"odinger equation with a radial $\frac{1}{r^2}$ potential. Such a potential requires boundary conditions at $r=0$ for the Hamiltonian to be self-adjoint; one needs to distinguish between several cases depending on the strength of the potential. Here we are in the ``most attractive'' case in which the coefficient $-(\frac{1}{4}+k^2)$ is less than $-\frac{1}{4}$. The boundary conditions and their solutions for all possible cases are discussed, e.g., in \cite{Narnhofer1974, *Kunstatter2009}. The solutions in our case are similar to these known ones; nevertheless we find it instructive to construct them explicitly. Readers not interested in the derivation may find the general form of physical wave functions with required boundary conditions in \eref{norm-wf}.

To see which are the allowed solutions we proceed in a case by case analysis. By considering two wave functions of purely positive $\lambda_1$ and $\lambda_2$,
\begin{eqnarray}
\fl\Psi_1(v,\varphi,t)&=& \int_{-\infty}^{\infty} \frac{\dd k}{2\pi} \int_{0}^{\infty}\frac{\dd \lambda}{2\pi\hbar}\, e^{i\frac{\lambda}{\hbar}t}e^{ik\varphi}\left( \alpha_1(k,\lambda)\JBessel{ik}{\sqrt{\lambda}}{v}+\beta_1(k,\lambda)\JBessel{-ik}{\sqrt{\lambda}}{v}\right) \nonumber\,,
\\\fl\Psi_2(v,\varphi,t)&=& \int_{-\infty}^{\infty} \frac{\dd k}{2\pi} \int_{0}^{\infty}\frac{\dd \lambda}{2\pi\hbar}\, e^{i\frac{\lambda}{\hbar}t}e^{ik\varphi}\left( \alpha_2(k,\lambda)\JBessel{ik}{\sqrt{\lambda}}{v}+\beta_2(k,\lambda)\JBessel{-ik}{\sqrt{\lambda}}{v}\right) \nonumber\,,
\end{eqnarray}
one can first show that there is no restriction on these wavefunctions coming from $v=\infty$: using the asymptotic form of the Bessel function for large arguments
\begin{equation}
J_\nu(z)\longrightarrow \sqrt{\frac{2}{\pi z}}\cos\left(z-\frac{\nu\pi}{2}-\frac{\pi}{4}\right), \hspace{3mm}{z\rightarrow \infty},
\end{equation}
one sees that all contributions to \eqref{bound} coming from $v=\infty$ are multiplied by terms of the form $\delta(\sqrt{\lambda_1}+\sqrt{\lambda}_2)(\lambda_1-\lambda_2)$ or $\delta(\sqrt{\lambda_1}-\sqrt{\lambda}_2)(\lambda_1-\lambda_2)$, and hence they all vanish.

Then, using the small argument asymptotic form of the Bessel function 
\begin{equation}
J_\nu(z)\longrightarrow \frac{e^{\nu\log\left(\frac{z}{2}\right)}}{\Gamma(\nu+1)}, \hspace{3mm}{z\rightarrow 0},
\label{small-arg-bessel}
\end{equation}
the boundary condition \eqref{bound} gives the following condition in the $v=0$ limit:
\begin{equation}
\overline{\alpha_1}(\lambda_1,k)\alpha_2(\lambda_2,k)e^{-ik\log\sqrt{\frac{\lambda_1}{\lambda_2}}}-\overline{\beta_1}(\lambda_1,k)\beta_2(\lambda_2,k)e^{ik\log\sqrt{\frac{\lambda_1}{\lambda_2}}}=0.
\end{equation}
The general solution to this condition is
\begin{equation}
\beta_I(k,\lambda)=\alpha_I(k,\lambda)e^{-2i\theta(k)}e^{ik\log\left(\frac{\lambda}{\lambda_0}\right)}\quad (I=1,2)
\end{equation}
where $\theta(k)$ is an arbitrary real function of $k$ and $\lambda_0>0$ is an arbitrary reference scale. A general parametrisation for such wave functions is then given by
\begin{equation}
\fl
\Psi_+(v,\varphi,t)=\int_{-\infty}^\infty \frac{\dd k}{2\pi} \int_0^\infty \frac{\dd \lambda}{2\pi\hbar} \ e^{i k\varphi}e^{i \lambda\frac{t}{\hbar}}A(k,\lambda)\,\Re\left[e^{i\theta(k)-ik\log\sqrt{\frac{\lambda}{\lambda_0}}} \JBessel{ik}{\sqrt{\lambda}}{v}\right]
\label{sols-bound-j}
\end{equation}
where $A(k,\lambda)$ is now an arbitrary function only constrained by the requirement that the state should be normalisable, and $\Re$ denotes taking the real part. Hence, we see that for $\lambda>0$ only certain real combinations of Bessel functions are allowed and that different self-adjoint extensions parametrised by the free function $\theta(k)$ are possible.

Now consider $\lambda<0$. At large argument the modified Bessel functions $I$ behave as
\begin{equation}
I_\nu(z)\longrightarrow \frac{e^{z}}{\sqrt{2\pi z}}, \hspace{3mm} z\rightarrow \infty,
\end{equation}
hence they are not integrable under the inner product \eqref{inner-prod-t}. Therefore we only consider the functions $K_{ik}\left(\frac{\sqrt{-\lambda}}{\hbar}v\right)$ and write a wave function with support only on $\lambda<0$ as
\begin{equation}
\Psi_-(v,\varphi,t)=\int_{-\infty}^\infty \frac{\dd k}{2\pi} \int_{-\infty}^0 \frac{\dd \lambda}{2\pi\hbar}\, e^{i\lambda \frac{t}{\hbar}} e^{ik\varphi}B(k,\lambda)\KBessel{ik}{\sqrt{-\lambda}}{v}
\label{general-neg-lambda}
\end{equation}
where $B(k,\lambda)$ is, at this point, an arbitrary complex function.
Inserting two wave functions of this form into \eqref{bound} and using the asymptotic form 
\begin{equation}
K_\nu(z)\longrightarrow \frac{1}{2}\left(\Gamma(-\nu)e^{\nu\log\left(\frac{z}{2}\right)}+\Gamma(\nu)e^{-\nu\log\left(\frac{z}{2}\right)}\right), \hspace{3mm} z\rightarrow 0,
\label{small-arg-bessel-k}
\end{equation}
we find that for any fixed $k$, two allowed negative values $\lambda_1$ and $\lambda_2$ for $\lambda$ have to satisfy
\begin{equation}
\log\left(\sqrt{\frac{\lambda_1}{\lambda_2}}\right)=\frac{\pi n}{k}
\end{equation}
where $n$ is integer. In other words, for given $k$ the only allowed $\lambda$ are of the form
\begin{equation}
\lambda_n^k=\lambda_G^k e^{-\frac{2\pi n}{k}}, \hspace{3mm} n\in \mathbb{Z}
\label{ln}
\end{equation}
for some $\lambda_G^k<0$. 
This requirement already restricts the form \eqref{general-neg-lambda} to
\begin{equation}
\Psi_-(v,\varphi,t)=\int_{-\infty}^\infty \frac{\dd k}{2\pi} \,e^{ik\varphi}\sum_{n=-\infty}^{\infty} e^{i\lambda_n^k \frac{t}{\hbar}} B(k,\lambda_n^k)\KBessel{ik}{\sqrt{-\lambda_n^k}}{v}.
\label{sols-bound-k}
\end{equation}

Evaluating the boundary condition \eqref{bound} with functions of positive $\lambda$ \eqref{sols-bound-j} and functions of negative $\lambda$ \eqref{sols-bound-k} (and using again the small argument asymptotic form for the Bessel functions) we finally find that $\lambda_G^k$ needs to be fixed to \footnote{$\lambda_0$ is an arbitrary positive number: changing $\lambda_0\rightarrow\lambda'_0,\;\theta(k)\rightarrow\theta(k)+k\log\sqrt{\frac{\lambda_0}{\lambda_0'}}$ leaves $\lambda_G^k$ invariant.}
\begin{equation}
\lambda_G^k=-\lambda_0e^{-\frac{\pi}{k}+\frac{2\theta(k)}{k}}.
\end{equation}
For negative values of $\lambda$, the set of allowed wave functions is rather severely restricted: the quantum theory exhibits a curious type of discreteness where for each wavenumber $k$ only a discrete set of ($k$-dependent) values of $\lambda$ are possible. 

Wave functions of the form \eqref{sols-bound-j} and \eqref{sols-bound-k} have time-independent norm, but we still need to normalise them. Let us start with \eqref{sols-bound-j} and consider functions of the form 
\begin{equation}
\psi_{k,\lambda}(v,\varphi)=\alpha(k,\lambda)e^{ik\varphi}\Re\left[e^{i\theta(k)-ik\log\sqrt{\frac{\lambda}{\lambda_0}}} \JBessel{ik}{\sqrt{\lambda}}{v}\right],\quad \lambda>0.
\label{psi-lk-pos}
\end{equation}
We would like to find $\alpha(k,\lambda)$ such that $\braket{\psi_{\lambda_1,k_1}}{\psi_{\lambda_2,k_2}}=\left( 2\pi\right)^2\hbar\,\delta(k_1-k_2)\delta(\lambda_1-\lambda_2)$, i.e., these wave functions form an orthonormal basis (in the improper sense of Dirac delta normalisation). We already know that the $\psi_{k,\lambda}$ are orthogonal for different values of $\lambda$ because they are eigenstates of the Hamiltonian with different eigenvalues, hence we only need to find the right constant of proportionality. The $\varphi$ part of the eigenstates is already normalised by $\int \dd \varphi \ e^{i(k_2-k_1)\varphi}=2\pi\delta(k_1-k_2)$. We can then focus on evaluating
\begin{eqnarray}
\fl
\int_0^\infty \dd v\ v\, \bar{\psi}_{\lambda_1,k}\psi_{\lambda_2,k}&=\int_0^\infty \dd v \   v\,\Bigl( \bar{\alpha}(k,\lambda_1)\alpha(k,\lambda_2)\times\nonumber \\
 &\Re\left[e^{i\theta(k)-ik\log\sqrt{\frac{\lambda_1}{\lambda_0}}} \JBessel{ik}{\sqrt{\lambda_1}}{v}\right]\Re\left[e^{i\theta(k)-ik\log\sqrt{\frac{\lambda_2}{\lambda_0}}} \JBessel{ik}{\sqrt{\lambda_2}}{v}\right]\Bigr)\nonumber\\
& = \frac{1}{4}\bar{\alpha}_1\alpha_2\int_0^\infty \dd v \ v \left[ e^{-2i\theta(k)+ik\log\sqrt{\frac{\lambda_1\lambda_2}{\lambda_0^2}}}\JBessel{-ik}{\sqrt{\lambda_1}}{v}\JBessel{-ik}{\sqrt{\lambda_2}}{v} \nonumber \right. \\
&\left. +e^{ik\log\sqrt{\frac{\lambda_1}{\lambda_2}}}\JBessel{-ik}{\sqrt{\lambda_1}}{v}\JBessel{ik}{\sqrt{\lambda_2}}{v} + \ \textrm{complex conjugate}\right],
\label{calc-2}
\end{eqnarray}
where $\alpha_i\equiv\alpha(k,\lambda_i)$. We already know that this integral is of the form
\begin{equation}
\int_0^\infty \dd v\ v\, \bar{\psi}_{\lambda_1,k}\psi_{\lambda_2,k} = f(\lambda_1)\delta(\lambda_1-\lambda_2),
\label{intf}
\end{equation}
i.e., it represents a distribution to be integrated over $\lambda_i$. To evaluate the integral in this distributional sense, any possible finite terms $g(\lambda_1)\delta_{\lambda_1,\lambda_2}$ can be ignored since they would integrate to zero. The lower limit of (\ref{calc-2}) gives such a finite contribution since the Bessel functions are bounded near 0 and the prefactor $e^{ik\log\sqrt{\frac{\lambda}{\lambda_0}}}$ cancels their infinite oscillations. We hence focus on the upper limit, and use the asymptotic form 
\begin{equation}
\JBessel{ik}{\sqrt{\lambda}}{v}\longrightarrow \sqrt{\frac{2\hbar}{\pi \sqrt{\lambda}v}}\cos\left(\frac{\sqrt{\lambda}}{\hbar}v-\frac{ik\pi}{2}-\frac{\pi}{4}\right), \hspace{3mm}v \rightarrow \infty.
\label{large-arg-asympt}
\end{equation}
to determine the function $f(\lambda_1)$ in (\ref{intf})\footnote{To confirm that using the asymptotic form is sufficient to obtain the correct result, notice that if functions $\gamma,\eta,G$ and $F$ are such that
$$\int_0^\infty \dd v\; \gamma(v)=(\lim_{v\rightarrow\infty}G(v))-G(0)\,,\quad \int_0^\infty \dd v\; \eta(v)=(\lim_{v\rightarrow\infty}F(v))-F(0),$$
the integrals are only defined in a distributional sense and $\lim_{v\rightarrow\infty}(\gamma(v)-\eta(v))=0$, we must have $\lim_{v\rightarrow\infty}(F'(v)-G'(v))=0$; then, if $F(0)$ and $G(0)$ are finite, the difference between the two integrals can only be a finite term and thus they give (in our context) the same result as distributions.}: we write the product of two Bessel functions as a combination of cosines and sines with argument $\left(\frac{\sqrt{\lambda_1}}{\hbar}\pm\frac{\sqrt{\lambda_2}}{\hbar}\right)v$. Integrating these can give delta distributions using the identity
\begin{equation}
\int_0^\infty \dd x\,\cos(\kappa x)=\pi\delta(\kappa).
\end{equation}
We also have $\delta\left(\frac{\sqrt{\lambda_1}}{\hbar}+\frac{\sqrt{\lambda_2}}{\hbar}\right)=0$ since $\lambda_1$ and $\lambda_2$ are positive; yet another simplification is that for any $F(\lambda_1,\lambda_2)$ we can set $F(\lambda_1,\lambda_2)\delta(\lambda_1-\lambda_2)=F(\lambda_1,\lambda_1)\delta(\lambda_1-\lambda_2)$ and further simplify the integrand. After substituting \eqref{large-arg-asympt} into \eqref{calc-2} we then find that
\begin{eqnarray}
\fl &&\int_0^\infty \dd v \ v \bar{\psi}_{\lambda_1,k}\psi_{\lambda_2,k}\nonumber
\\\fl &=& \frac{\hbar\bar{\alpha}_1\alpha_2}{2\pi\sqrt{\lambda_1}}\int_0^\infty \dd v \left( \cos\left(k\log\frac{\lambda_1}{\lambda_0}-2\theta(k)\right)+ \cosh(k\pi)\right)\cos\left(v\left(\frac{\sqrt{\lambda_1}}{\hbar}-\frac{\sqrt{\lambda_2}}{\hbar}\right)\right)\nonumber
\label{calc-3}
\\\fl&=& \frac{\hbar|\alpha_1|^2}{2\sqrt{\lambda_1}}\left( \cos\left(k\log\frac{\lambda_1}{\lambda_0}-2\theta(k)\right)+ \cosh(k\pi)\right)\delta\left(\frac{\sqrt{\lambda_1}}{\hbar}-\frac{\sqrt{\lambda_2}}{\hbar}\right)\nonumber
\\\fl
&=&\hbar^2|\alpha_1|^2\left(\cos\left(-2\theta(k)+k\log\frac{\lambda_1}{\lambda_0}\right)+\cosh(k\pi)\right)\delta(\lambda_1-\lambda_2).
\end{eqnarray}
Therefore, an orthonormal basis for the eigenstates of positive energy is given by
\begin{equation}
\psi_{k,\lambda}(v,\varphi)=\frac{\sqrt{2\pi} e^{ik\varphi}\Re\left[e^{i\theta(k)-ik\log\sqrt{\frac{\lambda}{\lambda_0}}} \JBessel{ik}{\sqrt{\lambda}}{v}\right]}{\sqrt{\hbar\cos\left(-2\theta(k)+k\log\frac{\lambda}{\lambda_0}\right)+\hbar\cosh(k\pi)}}\,.
\end{equation}
At large $v$ the eigenstates $\psi_{k,\lambda}$ have the form
\begin{eqnarray}
\psi_{k,\lambda}(v,\varphi)\propto  \frac{e^{ik\varphi}}{\sqrt{\sqrt{\lambda}v}}&\left[ e^{-\frac{k\pi}{2}}\cos\left(\frac{\sqrt{\lambda}}{\hbar}v-\frac{\pi}{4}-\theta(k)+k\log\sqrt{\frac{\lambda}{\lambda_0}}\right)\right.\nonumber \\ 
&+\left.e^{\frac{k\pi}{2}}\cos\left(\frac{\sqrt{\lambda}}{\hbar}v-\frac{\pi}{4}+\theta(k)-k\log\sqrt{\frac{\lambda}{\lambda_0}}\right)\right]
\end{eqnarray}
which is a combination of outgoing and incoming plane waves with phase difference $\Theta(k,\lambda)=\pm\frac{\pi}{2}+2\theta(k)-k\log\left(\frac{\lambda}{\lambda_0}\right)$. This difference can be viewed as a phase shift from scattering through the $v=0$ region, as discussed in more detail in \cite{Gryb2019, *Gryb2019_2, Gryb2019_1}.

The eigenstates of negative $\lambda$ are easier to normalise since for each $k$ there is only a discrete set of allowed $\lambda$ values; these states can hence be properly normalised. Consider
\begin{equation}
\phi_{\lambda_n^k,k}(v,\varphi)=\beta(k,\lambda_n^k)e^{ik\varphi}\KBessel{ik}{\sqrt{-\lambda_n^k}}{v}.
\label{phi-lk-neg}
\end{equation}
We would like to find $\beta(k,\lambda_n^k)$ such that $\braket{\phi_{\lambda_{n_1}^{k_1},k_1}}{\phi_{\lambda_{n_2}^{k_2},k_2}}=2\pi\delta(k_1-k_2)\delta_{n_1,n_2}$. Once again, because the $\varphi$ part of the eigenstates is already normalised, it is enough to analyse 
\begin{eqnarray}
\fl
\int_0^\infty \dd v \ v\, \bar{\phi}_{\lambda_{n_1}^k,k}\phi_{\lambda_{n_2}^k,k}=\bar{\beta}_1\beta_2\int_0^\infty \dd v \ v \KBessel{ik}{\sqrt{-\lambda_{n_1}^k}}{v}\KBessel{ik}{\sqrt{-\lambda_{n_2}^k}}{v}
\label{calc-4}
\end{eqnarray}
(recall that the modified Bessel functions $K$ are real). Here $\beta_i\equiv\beta(k,\lambda_{n_i}^k)$. For these integrals it is known (see p.~658, formula 6.521(3) in \cite{Integrals}) that
\begin{equation}
\int_0^\infty \dd x \ x\, K_\nu(ax)K_\nu(bx)=\frac{\pi(ab)^{-\nu}(a^{2\nu}-b^{2\nu})}{2\sin(\nu \pi )(a^2-b^2)}.
\label{calc-5}
\end{equation}
Hence, given the form of the eigenvalues \eqref{ln} and taking the limit $a\rightarrow b$ in \eqref{calc-5} we have that $\eqref{calc-4}$ vanishes for $n_1\neq n_2$, while for $n_1=n_2$
\begin{equation}
\int_0^\infty\dd v \ v\, \left|\phi_{\lambda_{n}^k,k}\right|^2=|\beta(k,\lambda_n^k)|^2\frac{\pi \hbar^2 k}{-2\lambda_n^k\sinh(k\pi)}.
\end{equation}
Therefore, the correct normalisation for negative energy eigenstates is
\begin{equation}
\phi_{\lambda_n^k,k}(v,\varphi)=\frac{1}{\hbar}\sqrt{\frac{-2\lambda_n^k\sinh(k\pi)}{\pi k}}e^{ik\varphi}\KBessel{ik}{\sqrt{-\lambda_n^k}}{v}.
\end{equation}

We can now finally write the general wave function of the universe in terms of the orthonormal basis $\lbrace \phi_{\lambda_n^k,k},\psi_{k,\lambda}\rbrace$ in the following way:
\begin{eqnarray}
\fl
\Psi(v,\varphi,t)=&\int_{-\infty}^\infty   \frac{\dd  k}{2\pi} \  e^{ik\varphi} \left[ \sum_{n=-\infty}^{\infty} B(k,\lambda_n^k)\frac{1}{\hbar} \sqrt{\frac{-2\lambda_n^k\sinh(k\pi)}{\pi k}}e^{i\lambda_n^k\frac{t}{\hbar}}\KBessel{ik}{\sqrt{-\lambda_n^k}}{v}\right. \nonumber \\
\fl 
 &+\left. \int_0^\infty \frac{\dd \lambda}{2\pi\hbar}\, e^{i\lambda\frac{t}{\hbar}}\, A(k,\lambda)\frac{\sqrt{2\pi}\Re\left[e^{i\theta(k)-ik\log\sqrt{\frac{\lambda}{\lambda_0}}} \JBessel{ik}{\sqrt{\lambda}}{v}\right]}{\sqrt{\hbar\cos\left(-2\theta(k)+k\log\frac{\lambda}{\lambda_0}\right)+\hbar\cosh(k\pi)}}\right],
 \label{norm-wf}
\end{eqnarray}
where $A$ and $B$ need to satisfy $\int_{-\infty}^\infty \frac{\dd k}{2\pi} \ \sum_{n=-\infty}^{\infty} |B(k,\lambda_n^k)|^2+\int_{-\infty}^\infty \frac{\dd k}{2\pi} \int_0^\infty \frac{\dd \lambda}{2\pi\hbar} \ |A(k,\lambda)|^2=1$. Such a wave function is guaranteed to have unit norm under the inner product \eqref{inner-prod-t}. We remind the reader that in \eref{norm-wf} $\theta(k)$ is a free function specifying the choice of self-adjoint extension, $\lambda_0>0$ is an arbitrary reference scale (arbitrary since the difference between any two such scales can be absorbed in $\theta(k)$) and $\lambda_n^k=-\lambda_0e^{-\frac{(2n+1)\pi}{k}+\frac{2\theta(k)}{k}}$ are the allowed negative $\lambda$ values.

\subsection{The volume clock $\log(v/v_0)$}
\label{v-clock}

In the last section we chose $t$ to be the clock variable and constructed a Hilbert space over the remaining variables. While this choice of clock was  justified by the fact that $t$ is a good classical clock, classically there are other possible choices; in particular one could take $\varphi$ or, in the case of $\lambda\ge 0$, the volume $v$. In this section we will choose $v$ -- or rather $\log(v/v_0)$ -- as a quantum clock, and compare the resulting theory to the Schr\"odinger quantum theory that was obtained for time $t$. 

 After multiplying by $v^2$, the Wheeler--DeWitt equation \eqref{wdw} can be rewritten as
\begin{equation}
\left(\hbar^2\left(v\pdv{}{v}\right)^2-\hbar^2\frac{\partial^2}{\partial \varphi^2}-i\hbar v^2\pdv{}{t}\right)\Psi(v,\varphi,t)=0.
\label{wdw-logv}
\end{equation}
This equation looks like a Klein--Gordon equation in the variables $(v,\varphi)$ where $\log(v/v_0)$ plays the r\^{o}le of time, since $v\pdv{}{v}=\pdv{}{\log(v/v_0)}$ (where $v_0$ is an arbitrary parameter with dimensions length$^{3/2}$ needed for dimensional reasons) and the last term corresponds to a $v$-dependent potential. This suggests defining the inner product
\begin{equation}
\fl
\braket{\Psi}{\Phi}_{KG}:=i\int_{-\infty}^\infty\dd t \int_{-\infty}^\infty \dd \varphi  \ \left[ \bar{\Psi}(v,\varphi,t)v\pdv{}{v}\Phi(v,\varphi,t)-\Phi(v,\varphi,t)v\pdv{}{v}\bar{\Psi}(v,\varphi,t)\right].
\end{equation}
This inner product is sesquilinear but might not be positive definite. Indeed, a general solution to \eqref{wdw-logv} of the form 
\begin{equation}
\fl
\Psi(v,\varphi,t)=\int_{-\infty}^\infty \hspace{-1mm}\frac{\dd k}{2\pi} \int_{-\infty}^\infty \hspace{-1mm}\frac{\dd \lambda}{2\pi\hbar}  e^{i\frac{\lambda}{\hbar}t}e^{ik\varphi}\hspace{-1mm}\left( \alpha(k,\lambda)\JBessel{i\abs{k}}{\sqrt{\lambda}}{v}+\beta(k,\lambda)\JBessel{-i\abs{k}}{\sqrt{\lambda}}{v}\right),
\label{gen-sol-wdw-3}
\end{equation} 
has norm squared
\begin{equation}
\norm{\Psi}^2_{KG}=\frac{2}{\pi}\int_{-\infty}^{\infty} \ddd{k} \int_{-\infty}^\infty \frac{\dd\lambda}{2\pi\hbar} \left[-\abs{\alpha(k,\lambda)}^2+\abs{\beta(k,\lambda)}^2 \right]\sinh( \abs{k}\pi),
\end{equation}
which is not positive in general. However, it is easy to redefine the Klein--Gordon inner product to obtain a positive definite inner product $\braket{\ }{\ }_{KG'}$. Notice that we can decompose the wave function \eqref{gen-sol-wdw-3} into positive and negative frequencies $\Psi=\Psi_++\Psi_-
$,
\begin{eqnarray}
\Psi_-=\int_{-\infty}^\infty\frac{\dd k}{2\pi} \int_{-\infty}^\infty \frac{\dd \lambda}{2\pi\hbar}  e^{i\frac{\lambda}{\hbar}t}e^{ik\varphi}\alpha(k,\lambda)\JBessel{i\abs{k}}{\sqrt{\lambda}}{v}, \nonumber \\
\Psi_+=\int_{-\infty}^\infty\frac{\dd k}{2\pi} \int_{-\infty}^\infty \frac{\dd \lambda}{2\pi\hbar}  e^{i\frac{\lambda}{\hbar}t}e^{ik\varphi}\beta(k,\lambda)\JBessel{-i\abs{k}}{\sqrt{\lambda}}{v},
\end{eqnarray}
such that $\norm{\Psi_-}_{KG}\leq0$, $\norm{\Psi_+}_{KG}\geq0$ and $\braket{\Psi_-}{\Psi_+}_{KG}=0$: positive and negative frequency sectors are decoupled, where ``frequency'' does not refer to an eigenvalue of $v\pdv{}{v}$ but to the sign in the inner product $\braket{\ }{\ }_{KG}$. It would be possible to build a consistent ``single-universe'' quantum theory from the positive frequency sector only, and no ``third quantisation'' \cite{McGuigan1988} in which $\Psi$ would be promoted to a quantum field is necessary.

Here we will instead consider both sectors and construct a positive definite inner product from $\braket{\ }{\ }_{KG}$. For this it is enough to redefine the inner product of the negative frequency modes by adding a minus sign, i.e.,
\begin{equation}
\norm{\Psi}_{KG'}=\norm{\Psi_+}_{KG}-\norm{\Psi_-}_{KG}
\end{equation}
or more explicitly
\begin{equation}
\norm{\Psi}^2_{KG'}=\frac{2}{\pi}\int_{-\infty}^{\infty} \ddd{k} \int_{-\infty}^\infty \frac{\dd\lambda}{2\pi\hbar} \left[\abs{\alpha(k,\lambda)}^2+\abs{\beta(k,\lambda)}^2 \right]\sinh( \abs{k}\pi).
\label{kgp-norm}
\end{equation}
A somewhat surprising property of this inner product is that it treats positive and negative $\lambda$ modes in exactly the same way, in spite of the fact that for $\lambda<0$ the mode functions diverge or fall off exponentially at large $v$ rather than oscillating as for $\lambda>0$. (Recall that our definition of Bessel functions for negative $\lambda$ was given in \eqref{besseldef}.)

Again, we require time independence of the inner product, $v\pdv{}{v}\braket{\Psi}{\Phi}_{KG'}=0$. This does not add any extra requirement for the Bessel functions. The condition
\begin{equation}
\int_{-\infty}^\infty \dd t \left[\bar{\Psi}\pdv{}{\varphi}\Phi-\Phi\pdv{}{\varphi}\bar{\Psi} \right]_{\varphi=-\infty}^{\varphi=\infty}=0
\label{bound-2}
\end{equation}
is satisfied by all solutions to the Wheeler--DeWitt equation, as can be seen by Fourier transforming in $\varphi$: the explicit expression \eqref{kgp-norm} is already manifestly independent of $v$. This is of course fundamentally different from what we found in the Schr\"odinger quantum theory in Section \ref{t-clock}. The Hilbert space now contains all regular functions of the form \eqref{gen-sol-wdw-3}, whereas using the $t$ clock only functions of the form \eqref{norm-wf} -- which are equal weight combinations of Bessel functions $J_{ik}$ and $J_{-ik}$ -- are permitted. A normalised wave function formed only out of positive frequency modes of the form
\begin{equation}
\fl
\Psi(v,\varphi, t)=\int_{-\infty}^\infty \frac{\dd\lambda}{2\pi\hbar} \int_{-\infty}^\infty \ddd{k}\sqrt{\frac{\pi}{2 \sinh(\abs{k}\pi)}}\ B(k,\lambda)e^{ik\varphi}e^{i\lambda\frac{t}{\hbar}}\JBessel{-i\abs{k}}{\sqrt{\lambda}}{v},
\label{norm-J}
\end{equation}
where $\int_{-\infty}^\infty \frac{\dd\lambda}{2\pi\hbar} \int_{-\infty}^\infty \ddd{k}\ \abs{B(k,\lambda)}^2=1$, will be used later to study singularity resolution.

In the discussion of Section \ref{t-clock} where we chose a Schr\"odinger time $t$, demanding conservation of inner products was equivalent to the requirement that the Hamiltonian be self-adjoint, and led to a boundary condition on wave functions. We have already seen that for a volume clock and Klein--Gordon-type inner product no boundary condition arises. This can be partially understood by observing that there is also no Hamiltonian, appearing as time evolution operator in a Schr\"odinger equation, that we could require to be self-adjoint. To see this let us define $u=\log(v/v_0)$ and rewrite the Wheeler--DeWitt equation \eqref{wdw-logv} in the form
\begin{equation}
-\hbar^2\frac{\partial^2}{\partial u^2}\Psi(v,\varphi,t)=\hat{\mathcal{O}}(u)\Psi(v,\varphi,t),
\label{wdw-u}
\end{equation}
with $\hat{\mathcal{O}}(u)\Psi:=\left(-\hbar^2\frac{\partial^2}{\partial\varphi^2}-i\hbar v_0^2 e^{2u}\pdv{}{t} \right)\Psi$. Unitarity of the Klein--Gordon quantum theory can be interpreted as a self-adjointness condition on $\hat{\O}$:
\begin{equation}
\fl
\pdv{}{u}{\braket{\Psi}{\Phi}_{KG'}}=0 \iff \int \dd t\,  \dd \varphi \left[\bar{\Psi}(u,\varphi,t)\hat{\O}\Phi(u,\varphi,t)-\Phi(u,\varphi,t)\hat{\O}\bar{\Psi}(u,\varphi,t) \right]=0.
\end{equation}
$\hat{\O}(u)$ is self-adjoint on $L^2(\mathbb{R}^2,\dd t \dd \varphi)$ for each value of $u$, since it is a combination of self-adjoint momentum operators in Schr\"odinger quantum mechanics. Hence, for wave functions that are square-integrable in the usual sense, we immediately have $\pdv{}{u}{\braket{\Psi}{\Phi}_{KG'}}=0$.  There is, however, no interpretation of $\hat{\O}$ as a Hamiltonian.

If $\hat{\O}$ was independent of $u$ and had positive eigenvalues, then the second order Klein--Gordon equation \eqref{wdw-u} could be replaced by two Schr\"odinger equations\footnote{More generally, one can rewrite Klein--Gordon-like equations as two coupled Schr\"odinger equations in matrix form \cite{Feshbach1958}. Various possible inner products in this general setting are discussed in \cite{Mostafazadeh2003}. There, one also finds that time-dependent Klein--Gordon operators generally do not lead to unitary dynamics.}
\begin{equation}
i\hbar \pdv{}{u}\Psi_\pm(u,\varphi,t)=\pm\sqrt{\hat{\O}}\,\Psi_\pm(u,\varphi,t).
\end{equation}
In this case, self-adjointness and positivity of $\hat{\O}$ ensure that $\sqrt{\hat{\O}}$ is a well-defined operator; $\sqrt{\hat{\O}}$ becomes a Hamiltonian and the general solution of the Klein--Gordon equation is of the form $\Psi=\Psi_++\Psi_-$ for two solutions of the two Schr\"odinger equations.

In our case however, $\hat{\O}$ depends on $u$. Here we would be looking for an associated Schr\"odinger equation of the form
\begin{equation}
i\hbar\pdv{}{u}\Psi(u,\varphi,t)= \hat{\H}\Psi(u,\varphi,t),
\label{schrodi}
\end{equation}
where agreement with \eqref{wdw-u} would demand that $\hat{\H}$ satisfies
\begin{equation}
\hat{\H}^2+ i\hbar\pdv{\hat{\H}}{u}=\hat{\O}.
\end{equation}
If we assume that both $\hat{\H}$ and $\hat{\O}$ are self-adjoint this equation does not have any solution, as can be seen from taking expectation values (again, for states for which both sides would be well-defined)
\begin{equation}
\expval{\hat{\H}^2}+ i\hbar\pdv{\expval{\hat{\H}}}{u}=\expval{\hat{\O}}.
\label{schr-kg-eqv}
\end{equation}
The right-hand side always takes real values whereas the left-hand side does not unless $\pdv{\hat{\H}}{u}=0$ which would imply that $\hat{\O}$ must also be time-independent.

Notice that already at the classical level, the constraint 
\begin{equation}
\mathcal{C}=-\pi_v^2 +\frac{\pi_\varphi^2}{v^2}+\lambda
\end{equation}
does not admit a splitting $\mathcal{C}=-\pi_v^2 + \mathcal{H}^2$ such that $\mathcal{H}^2$ would be a Dirac observable.

There is thus no obvious link of our quantum theory to a Schr\"odinger quantum theory with Hamiltonian that could be required to be self-adjoint. To the best of our knowledge, the closest interpretation in terms of an effective Schr\"odinger description is at the semiclassical level, where it requires a choice of complex Schr\"odinger time. We summarise this argument here as given in \cite{Bojowald2011}; see also \cite{Bojowald2010} for further discussion and various examples in which the effective Schr\"odinger time describing evolution of a constrained system acquires an imaginary part.

The idea is to relax the requirement that the Schr\"odinger equation \eqref{schrodi} is a differential equation in the same time variable $u$ as the second-order equation \eqref{wdw-u}. Instead one considers a general time variable $\tau$ such that $\pdv{}{\tau}=\pdv{}{u}$. Writing $\tau=u+\delta$ and Taylor expanding $\expval{\hat{\H}^2}$ to first order in $\delta$, the condition analogous to \eref{schr-kg-eqv} is now
\begin{equation}
\fl
\expval{\hat{\H}(u)^2}+\expval{\hat{\H}(u)\pdv{\hat{\H}(u)}{u}+\pdv{\hat{\H}(u)}{u}\hat{\H}(u)}\delta+ i\hbar\pdv{\expval{\hat{\H}(u)}}{u}=\expval{\hat{\O}(u)}.
\label{newcond}
\end{equation}
If we now focus on semiclassical states for which covariances are small and one can approximate $\expval{\hat{\H}(u)\pdv{\hat{\H}(u)}{u}}= \expval{\pdv{\hat{\H}(u)}{u}\hat{\H}(u)}= \expval{\hat{\H}(u)}\pdv{\expval{\hat{\H}(u)}}{u}$, \eref{newcond} can be solved if we set $\tau=u-\hbar \frac{i}{2\expval{\hat{\H}}}$ at first order of perturbation. Notice that this is a semiclassical expansion due to the appearance of $\hbar$. 

\begin{figure}[!h]
\centering
\begin{subfigure}{0.49\textwidth}
\centering
\includegraphics[scale=0.32]{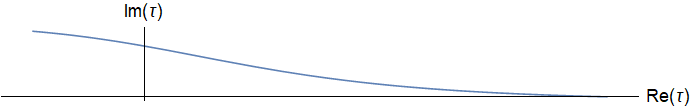}
\caption{Real and imaginary part of $\tau$ along a classical solution. The singularity corresponds to $\Re(\tau)=-\infty$.}
\end{subfigure}
\begin{subfigure}{0.49\textwidth}
\centering
\includegraphics[scale=0.34]{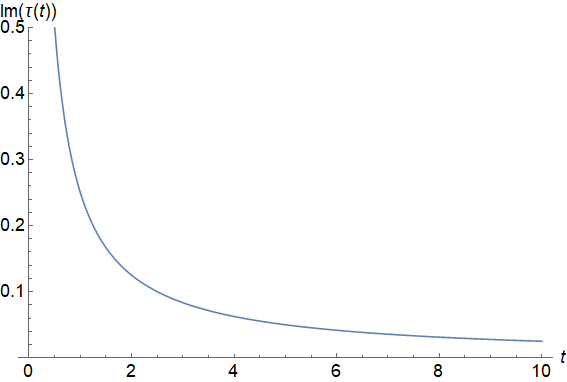}
\caption{Imaginary part of $\tau$ along the same classical solution parametrised by Schr\"odinger time $t$. The singularity is at $t=\frac{1}{2}$.}
\end{subfigure}
\caption{Effective complex Schr\"odinger time $\tau$ along a classical solution for $t_0=0$, $\lambda=\pi_\varphi=1$, $\hbar=1$. The behaviour is similar for other values of these parameters.}
\label{semiclass-t-plots}
\end{figure}

It is interesting to look at the real and imaginary parts of this complex Schr\"odinger time evaluated on a classical solution. Here we replace $\expval{\hat{\H}}\rightarrow v\pi_v$ where $v$ and $\pi_v$ follow a classical trajectory parametrised by
\begin{equation}
v(t)=\sqrt{-\frac{\pi_\varphi^2}{\lambda}+4\lambda(t-t_0)^2},\quad \pi_v(t)=\frac{2\lambda(t_0-t)}{\sqrt{-\frac{\pi_\varphi^2}{\lambda}+4\lambda(t-t_0)^2}}
\end{equation}
so that $v\pi_v=2\lambda(t_0-t)$. For an expanding solution with $\lambda>0$, $v\pi_v\rightarrow -|\pi_\varphi|$ at the singularity; the imaginary part of $\tau$ is positive and bounded by $\hbar/2|\pi_\varphi|$ (see Figure \ref{semiclass-t-plots}). The imaginary part remains very small for macroscopic $\pi_\varphi$ with $|\pi_\varphi|\gg\hbar$.

The imaginary part of $\tau$ is relevant near the classical singularity but tends to zero as $v$ tends to infinity. This confirms the general expectation that far away from the singularity the behaviour of our quantum theory is well described by standard Schr\"odinger quantum theory with evolution in a real time variable, whereas close to the singularity deviations from standard quantum theory become noticeable, ultimately leading to different physical predictions of the two theories. Interestingly, the quantum behaviour of the ``perfect bounce'' model in \cite{Gielen2016, *Gielen2017} was also captured by semiclassical {\em complex} trajectories in conformal time, leading to a picture in which the classical singularity is avoided in the complex plane similar to quantum tunnelling. It would be interesting to investigate the connection between these different ways in which a complex semiclassical time emerges in this quantum cosmology near classical singularities.

\section{Singularity resolution}
\label{sing-res-sect}

The possibility of resolving classical singularities has always been one of the strongest motivations for studying quantum gravity. In this section we will study the predictions of the two quantum theories we have constructed regarding resolution of the classical singularity. We will show that in the Schr\"odinger time theory, generic semiclassical states resolve the singularity, whereas it is possible to construct wave functions in the volume time theory that follow the classical solution all the way to the singularity. This illustrates the main result of our paper: the fundamental property of resolving classical singularities in this model depends on the choice of clock used for quantisation. 

In the Schr\"odinger quantum theory, all allowed wave functions contain combinations of Bessel functions of order $ik$ and $-ik$. The asymptotic expression of these Bessel functions for small arguments \eqref{small-arg-bessel} implies that $J_{ik}(z)\propto e^{ik \log\frac{z}{2}}$ when $z\rightarrow 0$. Hence, the wave  functions of the universe are composed of an equal weight combination of plane waves going into and out of the classical singularity. These states are in a sense very non-classical, as they are not associated to a unique classical trajectory. 

This feature is not reproduced in the volume time theory, where we can choose a wave function that is a combination of only incoming or outgoing waves. To illustrate the difference between the two theories, we choose in the volume time theory a wavepacket composed only of outgoing waves. Concretely, our chosen states have the form
\begin{equation}
\fl
\Psi_1(v,\varphi,t)=\int_{-\infty}^{\infty} \ddd{k}\int_0^\infty \frac{\dd \lambda}{2\pi\hbar}\, A(k,\lambda) \frac{\sqrt{2\pi}\, e^{i\lambda\frac{t}{\hbar}}e^{ik\varphi}\,\Re\left[e^{i\theta(k)-ik\log\sqrt{\frac{\lambda}{\lambda_0}}} \JBessel{ik}{\sqrt{\lambda}}{v}\right]}{\sqrt{\hbar\cos\left(k\log\frac{\lambda}{\lambda_0}-2\theta(k)\right)+\hbar\cosh(k\pi)}}
\label{psi-1}
\end{equation}
and
\begin{equation}
\fl
\Psi_2(v,\varphi,t)=\int_{-\infty}^{\infty}\ddd{k} \int_{0}^\infty \frac{\dd{\lambda}}{2\pi\hbar}\,B(k,\lambda)  \sqrt{\frac{\pi}{2 \sinh(\abs{k}\pi)}}\ e^{ik\varphi}e^{i\lambda\frac{t}{\hbar}}\JBessel{-i\abs{k}}{\sqrt{\lambda}}{v},
\label{psi-2}
\end{equation}
where $A(k,\lambda)$ and $B(k,\lambda)$ are normalised: $\int \ddd{k} \frac{\dd \lambda}{2\pi\hbar} \abs{A(k,\lambda)}^2=\int \ddd{k} \frac{\dd \lambda}{2\pi\hbar} \abs{B(k,\lambda)}^2=1$. We focus on positive $\lambda$ because it is easier to build semiclassical states for $\Psi_1$ as all values of $k$ and $\lambda>0$ are allowed. We stress that these states live in different Hilbert spaces: $\Psi_1$ is in the Hilbert space of the Schr\"odinger time theory while $\Psi_2$ defines a state in the volume time theory. We can then compare expectation values $\expval{v(t)}_{\Psi_1}$ in the Schr\"odinger time theory and $\expval{t(v)}_{\Psi_2}$ in the volume time theory. These expectation values are the quantum analogues of the classical Dirac observables \eqref{voft} and \eqref{tofv}. 

\subsection{Results for $\expval{t(v)}_{\Psi_2}$}

It is possible to obtain analytical results for $\expval{t(v)}$. Indeed, we have
\begin{eqnarray}
\fl
\expval{t(v)}_{\Psi_2} &
\hspace*{-12mm}= \left<\Psi_2\right|t\left|\Psi_2\right> \nonumber\\
&\hspace*{-12mm}=i \frac{\pi v}{2}\int \frac{\dd \lambda_1\,\dd \lambda_2}{(2\pi\hbar)^2}\,  \ddd{k} \,\dd t \,t\,  e^{-i(\lambda_1-\lambda_2)\frac{t}{\hbar}}\frac{\bar{B}(k,\lambda_1)B(k,\lambda_2)}{\sinh(\abs{k}\pi)}\times \nonumber \\
\fl&\hspace*{-6mm}\left[\JBessel{i\abs{k}}{\sqrt{\lambda_1}}{v}\partial_v\JBessel{-i\abs{k}}{\sqrt{\lambda_2}}{v}-\JBessel{-i\abs{k}}{\sqrt{\lambda_2}}{v}\partial_v\JBessel{i\abs{k}}{\sqrt{\lambda_1}}{v} \right]
\end{eqnarray}
and the $t$ integral can be done explicitly using the general expression
\begin{eqnarray}
&&\int \frac{\dd \lambda_1\,\dd \lambda_2}{(2\pi\hbar)^2}\, \dd t \, t\,  e^{-i(\lambda_1-\lambda_2)\frac{t}{\hbar}}F(\lambda_1,\lambda_2)\nonumber
\\&=&-\frac{i\hbar}{2}\int \frac{\dd \lambda}{2\pi\hbar}\left(\frac{\partial F(\lambda_1,\lambda_2)}{\partial\lambda_1}-\frac{\partial F(\lambda_1,\lambda_2)}{\partial\lambda_2}\right)\Big|_{\lambda_1=\lambda_2=\lambda}
\end{eqnarray}
leading to
\begin{eqnarray}
\expval{t(v)}_{\Psi_2}=&\int\frac{\dd \lambda}{2\pi\hbar}\ddd{k}\abs{B(k,\lambda)}^2 f(v,k,\lambda) \nonumber\\
&+\frac{i\hbar}{2}\int \frac{\dd \lambda}{2\pi \hbar}\ddd{k}\left[ \bar{B}(k,\lambda)\partial_\lambda B(k,\lambda)-\partial_\lambda\bar{B}(k,\lambda)B(k,\lambda)\right]
\label{expvalt}
\end{eqnarray}
where 
\begin{eqnarray}
\fl
f(v,k,\lambda)&=&\frac{\pi}{4\hbar \sinh(\abs{k}\pi)}\left[ v^2\abs{\JBessel{1+i\abs{k}}{\sqrt{\lambda}}{v}}^2+ \left(\frac{2\hbar^2k^2}{\lambda}+v^2\right)\abs{\JBessel{i\abs{k}}{\sqrt{\lambda}}{v}}^2 \right.  \\
\fl
&&-\left. \frac{i \hbar \abs{k} v}{\sqrt{\lambda}}\left[ \JBessel{i\abs{k}}{\sqrt{\lambda}}{v}\JBessel{-1-i\abs{k}}{\sqrt{\lambda}}{v}-\textrm{c.c.}\right]\right].\nonumber
\label{f}
\end{eqnarray}
We would like to compare \eqref{expvalt} with the behaviour of classical solutions
\begin{equation}
t_c(v)=\frac{1}{2}\sqrt{\frac{\hbar^2k_c^2}{\lambda_c^2}+\frac{v^2}{\lambda_c}},
\label{class-sol}
\end{equation}
where $\hbar k_c$ is the classical value of $\pi_\varphi$ and $\lambda_c$ the classical value of the cosmological constant. To do so we need to build semiclassical states, which are not too widely spread around the central values $k_c$ and $\lambda_c$. We assume the form $B(k,\lambda)=\kappa(k)\chi(\lambda)$ where $\kappa(k)$ is extremely sharply peaked, so that we take $\kappa(k)f(k)\approx\kappa(k)f(k_c)$ for any function $f(k)$ while $\int \ddd{k} \abs{\kappa(k)}^2=1$. This will simplify the numerics as we will not integrate over $k$. We then choose $\chi(\lambda)$ to be a normalised Gaussian of mean $\lambda_c$,
\begin{equation}
\chi(\lambda)=C\sqrt{\frac{2\hbar\sqrt{\pi}}{\sigma}}e^{-\frac{(\lambda-\lambda_c)^2}{2\sigma^2}}.
\label{gaussian}
\end{equation}
The condition for the state to be normalised is $\int_{0}^\infty \frac{\dd\lambda}{2\pi \hbar}\abs{\chi(\lambda)}^2= 1$. As we are integrating only over positive $\lambda$, we need to add a constant of normalisation $C(\lambda_c,\sigma)$. We mostly use states for which $\frac{\sigma}{\lambda_c}$ is small; $C$ is then essentially equal to 1. The standard deviation $\sigma$ controls ``how quantum'' the state is: if $\sigma\ll \lambda_c$ the Gaussian is very peaked around the classical value of the cosmological constant, whereas for $\sigma\gtrsim \lambda_c$ there is significant quantum spreading. For $\sigma\ll \lambda_c$ we can perform a further approximation of \eqref{expvalt},
\begin{equation}
\expval{t(v)}_{\Psi_2}\approx \int\frac{\dd \lambda}{2\pi\hbar} \abs{\chi(\lambda)}^2 f(v,k_c,\lambda)\approx f(v,k_c,\lambda_c).
\label{approx-tv}
\end{equation}
Notice that our amplitude function $B(k,\lambda)$ is chosen to be real meaning there is no contribution from the second line in \eqref{expvalt}.
Expanding both the classical solution \eqref{class-sol} and the quantum expectation value  \eqref{approx-tv} around $v=0$, we find 
\begin{eqnarray}
\fl
t_c(v) & = \frac{\hbar |k_c|}{2\lambda_c}+\frac{v^2}{4\hbar |k_c|}-\frac{\lambda_cv^4}{16 \hbar^3|k_c|^3}+\frac{\lambda_c^2v^6}{32\hbar^5 |k_c|^5}+O\left(v^8\right) \label{tayexpclass},\\
\fl
f(v,k_c,\lambda_c)& = \frac{\hbar |k_c|}{2\lambda_c}+\frac{v^2}{4\hbar |k_c|}-\frac{\lambda_cv^4}{16\hbar^3(|k_c|+|k_c|^3)}+\frac{\lambda_c^2v^6}{32\hbar^5 (4|k_c|+5|k_c|^3+|k_c|^5)}+O\left(v^8\right).\nonumber
\end{eqnarray}
There is very close agreement, with the first difference only coming in at $O(v^4)$. We also see that $k_c$ (which is dimensionless) can be used as another measure of semi-classicality: $f(v,k_c,\lambda_c)$ and $t_c(v)$ agree in the limit of very large $|k_c|$. It is therefore insightful to study solutions with different values of $k_c$. 

\begin{figure}[!h]
\centering
\includegraphics[scale=0.5]{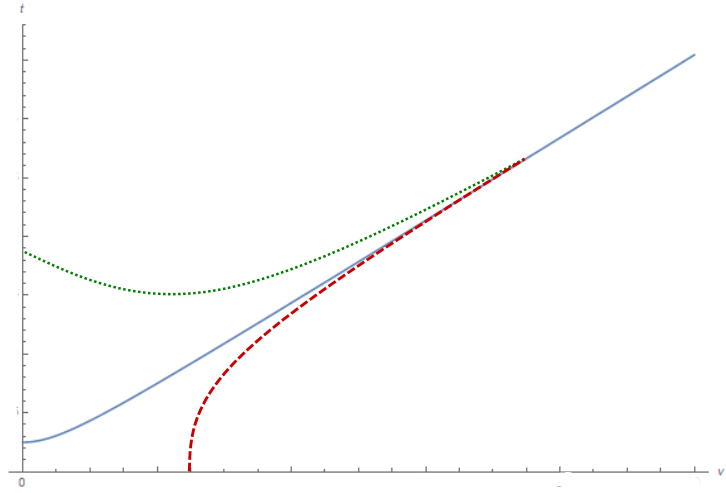}
\caption{Possible singularity resolution in volume time. The blue line corresponds to the classical solution $t(v)$. The green (dotted) line corresponds to a trajectory where $\expval{t(v)}$ starts moving backwards. The red (dashed) curve corresponds to a $\expval{t(v)}$ that becomes ill-defined. We would consider these scenarios as resolving the singularity.}
\label{sing}
\end{figure}

The theory shows no signs of departing from the classically singular behaviour. In order to claim singularity resolution one might require that the expectation value $\expval{t(v)}_\Psi$ become ill-defined as $|\partial_v\expval{t(v)}_\Psi|\rightarrow \infty$ for some value of $v$, in order to prevent the dynamics from reaching $v=0$. A weaker requirement would be departure from the classical property that $t$ is a globally monotonic function of $v$: if the clock $t$ started moving backwards ($\partial_v\expval{t(v)}_\Psi=0$ somewhere) this would also imply nonclassical and hence potentially nonsingular behaviour in our quantum theory. In this case, one could reach $v=0$ but the measurement of time using the $t$ clock would break down. These hypothetical scenarios are visualised in Figure \ref{sing}. 

\begin{figure}[!h]
\centering
\includegraphics[scale=0.5]{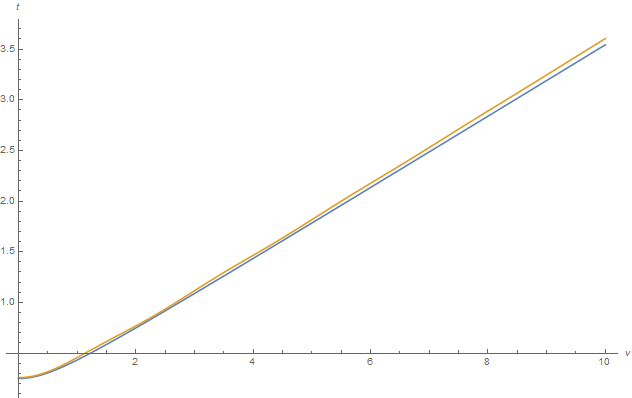}
\caption{Classical solution $t(v)$ (blue curve) and quantum expectation value $\expval{t(v)}_{\Psi_2}$ (yellow curve) for $\lambda_c=2$, $k_c=1$ and $\sigma=0.5$, in units $\lambda_0=\hbar=1$. Other values of the parameters do not alter significantly the form of the two curves.}
\label{tv-nicer}
\end{figure}

None of these scenarios occur here as the two lines of \eqref{tayexpclass} start to disagree only at order $v^4$. This can be confirmed by analysing Figure \ref{tv-nicer}, where we see that (for a particular choice of Gaussian state) the quantum expectation value does not deviate from the classical theory close to the singularity, but instead follows it very closely: states of the form $\Psi_2$ do not resolve the singularity. 

To quantify the difference between the classical solution and quantum expectation value seen in the graph, we introduce the relative difference between the quantum and the classical solution,
\begin{equation}
\Delta_{rel}(v)=\left(1-\frac{\expval{t(v)}_{\Psi_2}}{t_c(v)}\right).
\end{equation}
Some values for this difference are summarised in Table \ref{table1}. We observe that as $v$ increases, the relative difference (slowly) approaches 0. The negative sign indicates that the quantum expectation value is always greater than the classical solution. Of course, \eqref{tayexpclass} shows that there is a discrepancy between the two even for infinitely peaked states. As one might have expected, for the same values of $v$, $\lambda_c$ and $k_c$ the relative difference is  closer to zero for smaller $\sigma$, and already less peaked Gaussians ($\sigma=0.5$ and $\lambda_c=2$) give close agreement between quantum expectation value and classical solution. Increasing $k_c$ seems to slow the speed of convergence to the classical solution.

While the chosen states have $\lambda>0$, the same analysis can be performed for $\lambda<0$ with similar results; singularity resolution does not depend on the sign of $\lambda$ in the volume time theory which treats positive and negative values of $\lambda$ in the same way.

\begin{table}[h!]
\begin{center}
\begin{tabular}{c|c|c|c|c|c|c}
$\lambda_c$ & $k_c$ & $\sigma$ & $\Delta_{rel}(0.5)$ (\%) & $\Delta_{rel}(5)$ (\%) & $\Delta_{rel}(10)$ (\%) & $\Delta_{rel}(50)$ (\%)\\
\hhline{=|=|=|=|=|=|=} 
2 & 1 & 0.5 & -3.69 & -1.86 & -1.74& -1.65 \\
2 & 1 & 0.3 & -1.80 & -0.95 & -0.97 & -0.81 \\
2 & 1 & 0.1 & -0.95 & -0.56 & -0.75 & -0.43 \\
2 & 10 & 0.5  & -3.47 & -2.86 & -2.13& -1.32\\
10 & 1 & 1 & -4.16 & -0.47 & -0.56 & -0.56 \\
\end{tabular}
\caption{Relative difference for different values of the parameters $\lambda_c$, $k_c$ and $\sigma$. Despite some oscillations the two curves are slowly converging.}
\label{table1}
\end{center}
\end{table}

\subsection{Results for $\expval{v(t)}_{\Psi_1}$}

In the Schr\"odinger time theory, we are interested in the expectation value $\expval{v(t)}_{\Psi_1}$ of the volume as a function of the clock coordinate $t$. Expressions for inner products and expectation values are much harder to calculate analytically in this theory (for a discussion of analytical approximations in different limiting cases, see \cite{Gryb2019_1}), and as a result we will have to rely on numerics. Concretely, with our choice of state \eqref{psi-1} we are interested in computing 
\begin{eqnarray}
\fl
\expval{v(t)}_{\Psi_1} &
\hspace*{-12mm}= \left<\Psi_1\right|v\left|\Psi_1\right> \nonumber\\
&\hspace*{-12mm}= \int_0^\infty \dd v \; v^2\int \frac{\dd\lambda_1\,\dd\lambda_2}{(2\pi\hbar)^2}\,\ddd{k} e^{-i(\lambda_1-\lambda_2)\frac{t}{\hbar}}\bar{A}(k,\lambda_1) A(k,\lambda_2) \times\nonumber
\\& \frac{2\pi\,\Re\left[e^{-ik\log\sqrt{\frac{\lambda_1}{\lambda_0}}} \JBessel{ik}{\sqrt{\lambda_1}}{v}\right]\Re\left[e^{-ik\log\sqrt{\frac{\lambda_2}{\lambda_0}}} \JBessel{ik}{\sqrt{\lambda_2}}{v}\right]}{\sqrt{\hbar\cos\left(k\log\frac{\lambda_1}{\lambda_0}\right)+\hbar\cosh(k\pi)}\sqrt{\hbar\cos\left(k\log\frac{\lambda_2}{\lambda_0}\right)+\hbar\cosh(k\pi)}}
\label{vt-expval}
\end{eqnarray}
where we have chosen $\theta(k)=0$ for the self-adjoint extension of the theory. The choice of $\lambda_0$ is a choice of units in which the energy parameter $\lambda$ is measured. For the numerics we will set $\lambda_0=1$.  As in the previous analysis, we consider wave packets for which $A(k,\lambda)=\kappa(k)\chi(\lambda)$ with $\kappa(k)$ extremely sharply peaked around a wavenumber $k_c$ (so that the $k$ integral does not need to be evaluated numerically) and where $\chi(\lambda)$ is a Gaussian of the form \eqref{gaussian}. As the amplitude function $A(k,\lambda)$ is chosen to be real, $\expval{v(t)}_{\Psi_1}$ is symmetric with respect to time reversal $t\rightarrow -t$. Numerical evaluation of \eqref{vt-expval} then involves integrating over $\lambda_1,\lambda_2$ and $v$. The $v$ integral is a challenge as the Bessel functions oscillate rapidly near infinity. As a first consistency check, we have verified numerically that $\Psi_1$ is normalised, $\langle\Psi_1|\Psi_1\rangle=1$ to very high precision.

Before presenting our numerical results, let us formulate some general expectations based on the properties of the Schr\"odinger time quantum theory. For the classical theory, we saw in Section \ref{class-mod} that every classical solution (with nonvanishing scalar field energy, $\pi_\varphi\neq 0$) encounters a singularity at finite proper time: $v(t_{sing})=0$ for some finite $t_{sing}$. The classical evolution terminates there; at least for $\lambda\neq 0$, $v(t)$ cannot be continued beyond $t=t_{sing}$. However, our quantum theory is by construction unitary and so the time evolution of an initial state is well-defined along the entire $t$ axis. In this sense, there can be no singularity in the quantum theory as was argued already, e.g., in \cite{Gryb2019, *Gryb2019_2, Gotay1984}. At any point $t_p$ inside the range of the time coordinate $t$, we have a regular quantum state and it then follows that $\langle v(t_p)\rangle >0$ since there are no states in the Hilbert space with $\langle v(t_p)\rangle =0$. It is {\em a priori} still possible that $\langle v(t_p)\rangle =\infty$ somewhere, or that $\langle v(t)\rangle\rightarrow 0$ as $t\rightarrow\pm\infty$, and both of these behaviours might still be seen as singularities in the quantum theory. 

More generally, choosing a dynamical variable as clock means that this variable is treated as a parameter for the evolution of the other quantum variables. The fact that the volume time theory does not generically resolve the singularity could be seen as due to this choice, which implies the absence of explicit quantum fluctuations in the volume. In the Schr\"odinger time theory the time $t$ becomes an evolution parameter, which respect to which the volume has quantum fluctuations. The classical inevitability of reaching $v(t)=0$ can then be avoided.

In analogy with our discussion in the volume time theory, we would now say that any state $\Psi$ for which there is a $C_\Psi >0$ such that $\expval{v(t)}_\Psi\geq C_\Psi$ for all $t$ resolves the singularity. Indeed, this implies strong deviations from classical solutions in regions where the classical solution approaches zero. We would then also expect another strong departure from the classical theory, namely the existence of a point $t_0$ at which $\partial_t\expval{v(t)}_\Psi=0$. This would imply the relation $v(t)$ is no longer monotonic, and indicate the presence of a minimum (and bounce) for $\expval{v(t)}_\Psi$. To see all this explicitly for the state $\Psi_1$ as defined in \eqref{psi-1}, we now compare $\expval{v(t)}_{\Psi_1}$ with the classical solution
\begin{equation}
v_c(t)=\sqrt{4\lambda_c t^2- \frac{\hbar^2 k_c^2}{\lambda_c}}
\end{equation}
where $\lambda_c$ and $k_c$ correspond to the peak of the amplitude function $A(k,\lambda)$.

\begin{figure}[!h]
\centering
\includegraphics[scale=0.5]{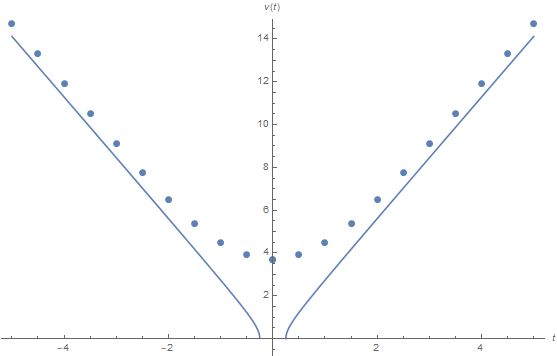}
\caption{Classical solution $v(t)$ (blue curve) and quantum expectation value $\expval{v(t)}_{\Psi_1}$ (blue dots) for values $\lambda_c=2$, $k_c=1$ and $\sigma=0.5$, in units $\lambda_0=\hbar=1$.} 
\label{vt-l2-k1}
\end{figure}
Figure \ref{vt-l2-k1} confirms our expectations, and shows the clear difference between classical and quantum solutions: $\expval{v(t)}_{\Psi_1}$ has a minimum clearly strictly above zero. As for the volume time theory, the quantum expectation value is always above the classical one and is converging slowly at late times. We also see that, unlike $v(t)$ which is only well-defined before the Big Crunch and after the Big Bang singularities, the quantum expectation value is  well-defined everywhere, going smoothly from the contracting to the expanding phase of the universe. This is exactly the behaviour of a nonsingular, bouncing universe, as was already observed in \cite{Gryb2019, *Gryb2019_2}.

\begin{figure}[!h]
\centering
\includegraphics[scale=0.5]{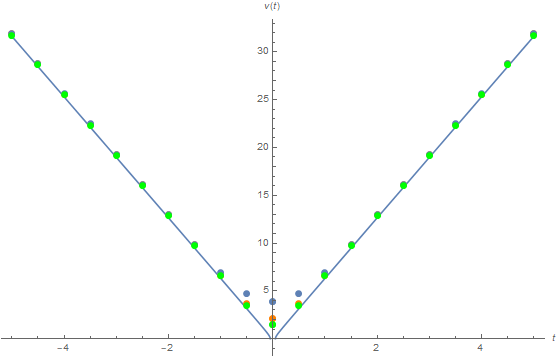}
\caption{Classical (blue line) and quantum solutions (dots) for different values of $\sigma$ ($\sigma=1$: blue, $\sigma=2$: orange, $\sigma=3$: green), with $\lambda_c=10$, $k_c=1$, and $\lambda_0=\hbar=1$.}
\label{vt-diff-sigma}
\end{figure}

Figure \ref{vt-diff-sigma} shows quantum solutions for different values of the standard deviation $\sigma$ in the Gaussian \eqref{gaussian}. Perhaps surprisingly, increasing the standard deviation reduces the difference between quantum expectation value and the classical solution near the classical singularity. 
In other words, states with larger quantum spread have a more abrupt transition between the two classical branches. This observation is related to the fact that the minimum value that $\expval{v(t)}_{\Psi_1}$ takes at $t=0$, which is in general a function of all free parameters $\lambda_c, k_c$ and $\sigma$, appears to decrease with increasing $\sigma$. This minimum value has no classical analogue, and therefore such behaviour would not be in conflict with the expectation that a semiclassical limit is $\sigma\rightarrow 0$. 

The analysis of \cite{Gryb2019_1} found the same general behaviour for the minimum value of $\expval{v(t)}$ and, in a limit in which the contribution of the cosmological constant dominates over the scalar field, found that analytically $\expval{v(0)}\propto 1/\sigma$ as a function of the standard deviation. Our numerical results (as shown in Figure \ref{vt-diff-sigma}) do not assume this limit and deviate from an exact relation  $\expval{v(0)}\propto 1/\sigma$ (at fixed $\lambda_c$ and $k_c$), but not very strongly: 
\begin{equation}
\fl
v(0)|_{\sigma=1}=3.90\,\pm\, 0.11,\quad v(0)|_{\sigma=2}=2.08\,\pm\, 0.04,\quad v(0)|_{\sigma=3}=1.47\,\pm\,0.02,
\end{equation} 
where the errors are numerical integration errors estimated by {\em Mathematica}.

The fact that larger $\sigma$ leads to a smaller minimum value for the volume can also be interpreted by noticing that $\lambda$ is conjugate to time and in some sense conjugate to $v$ (recall that $\JBessel{ik}{\sqrt{\lambda}}{v}\propto e^{ik \log\frac{\sqrt{\lambda}v}{2\hbar}}$ near $v=0$), so that it is the smaller spread in $t$ and $v$ which seems to bring the quantum expectation value closer to the classical solution.

As in the previous section, the quantum expectation value is always above the classical value but both are slowly converging as we evolve for longer times, as shown in Table \ref{table2}. The relative difference is again defined by
\begin{equation}
\Delta_{rel}(t)=\left(1-\frac{\expval{v(t)}_{\Psi_1}}{v_c(t)}\right).
\end{equation}
\begin{table}[!h]
\begin{center}
\begin{tabular}{c|c|c|c|c|c|c}
$\lambda_c$ & $k_c$ & $\sigma$ &   $\Delta_{rel}(5)$ (\%) & $\Delta_{rel}(10)$ (\%) & $\Delta_{rel}(15)$ (\%) & $\Delta_{rel}(20)$ (\%) \\
\hhline{=|=|=|=|=|=|=} 
2 & 1 & 0.5  & $-4.20$ $\pm$ 0.54 & $-1.01$ $\pm$ 0.17  & $-1.10$ $\pm$ 0.04 & $-0.82$ $\pm$ 0.01\\
10 & 1 & 1  &$-1.01$ $\pm$ 0.05 & $-0.47$ $\pm$ 0.02 &$-0.31$ $\pm$ 3.69 & $-0.20$ $\pm$ 0.01\\
10 & 1 & 2  &$-0.81$ $\pm$ 0.14 &$-0.36$ $\pm$ 9.84  & $0.14$ $\pm$ 27& $-0.01$ $\pm$ 0.14\\
10 & 1 & 3  &$-0.12$ $\pm$ 16 & 0.95 $\pm$ 25&0.23 $\pm$ 0.27 & 0.54 $\pm$ 1.96\\
10 & 3 & 1 &$-2.93$ $\pm$ 2.89 & $-1.45$ $\pm$ 0.02& $-0.94$ $\pm$ 0.01& $-0.69$ $\pm$ 0.03 \\
\end{tabular}
\caption{Relative difference between classical solution and quantum expectation value for different values of $\lambda_c$, $k_c$ and $\sigma$. The relative difference generally decreases over time. Integration error estimates by {\em Mathematica} were added to quantify the accuracy of the results. It appears that some high estimates are too high and the data are reliable.}
\label{table2}
\end{center}
\end{table}
Table \ref{table2} gives some more details about the behaviour of our numerical results. First of all, the relative difference  $\Delta_{rel}$ approaches zero over time, and generally the quantum solution is above the classical one: in the few cases where $\Delta_{rel}>0$, numerical errors are compatible with $\Delta_{rel}<0$. Given the same values of $\lambda_c$ and $k_c$, smaller values of $\sigma$ seem to give smaller numerical errors. As in the previous section, greater $k_c$ seems to slow down the convergence of quantum expectation values and classical solution. In general, some error estimates are very large, coming from the fact that the Bessel functions and the complex exponential can oscillate very rapidly. For these data points with very large error estimates, we have found that slightly changing parameters such as $\sigma$ or the time $t$ at which the integral is evaluated generically leads to a similar result which much smaller error estimate, suggesting that these error estimates significantly overestimate the actual numerical error.
Overall, the results are accurate enough to confirm the general behaviour of the Schr\"odinger time theory, in particular the apparently generic singularity resolution in this theory, in numerical examples. We again refer the reader to \cite{Gryb2019, *Gryb2019_2, Gryb2019_1} for many more details and a deeper numerical analysis of this theory.

\section{Conclusions}
\label{conclusio}

Singularity resolution in quantum cosmological models is a topic of ongoing research. A particularly pressing issue is the problem of time, which implies that quantum theories defined with respect to different clocks may not agree, and in particular make different predictions regarding resolution of the classical singularity. In this paper we contribute to this discussion by analysing flat FLRW universes filled with a scalar field and a matter component that can be interpreted as radiation, a cosmological constant as in unimodular gravity, or a more general perfect fluid, e.g., representing dust. As there is no preferred choice of time variable, we chose different dynamical variables to play the r\^ole of clock: a ``Schr\"odinger time'' $t$ representing one of the variables of the perfect fluid, which would correspond to unimodular time for dark energy and to conformal time for radiation, and a ``volume time'' $\log(v/v_0)$ where $v$ is the three-dimensional volume. 

These two choices of time coordinate were analysed by Gotay and Demaret \cite{Gotay1984} in a simpler but related model. In their terminology, the Schr\"odinger time $t$ is a ``slow'' clock: its domain is infinite but it only records a finite amount of time until the singularity is reached. The volume time on the other hand is ``fast'': the singularity is at infinity with respect to the clock $\log(v/v_0)$, or more generally at the boundary of its domain (which would be $v=0$ written in terms of $v$). Our results here confirm the conjecture given in \cite{Gotay1984} that fast clocks do not resolve the singularity whereas slow clocks lead to singularity-free theories. More precisely, the slow clock $t$ leads to a unitary and singularity-free theory, as shown in \cite{Gryb2019, *Gryb2019_2}; the price for this is boundary conditions which restrict the physical Hilbert space to nonclassical modes, each of which is an equal superposition of ingoing and outgoing. On the other hand, in the fast clock $\log(v/v_0)$ it is possible to construct states that follow the classical solution up to arbitrary accuracy. This confirms the picture of a well-defined quantum theory in which one can transition ``through'' the singularity, as given in \cite{Gielen2016}. The volume time theory does not have a self-adjoint Hamiltonian and, in its semiclassical limit, can be interpreted in terms of a complex Schr\"odinger time \cite{Bojowald2011}. We analysed the space of allowed wave functions in depth and presented both numerical and analytical results for singularity resolution.

Our results illustrate yet another example in which general covariance, one of the most praised characteristics of general relativity, is not maintained after quantisation as different clock choices lead to very different dynamics. 
Other examples are known \cite{Malkiewicz2019, *Malkiewicz2016, Bojowald2016}. One might conclude that deparametrised quantum theories, in which a time parameter is picked before quantisation, are thus subject to unacceptable ambiguities and that a more promising route would be to implement the programme of Dirac quantisation \cite{diracbook, Tate1992, Marolf2000} where an inner product can be constructed systematically, e.g., through group averaging. Dirac quantisation can be seen as implementing a clock-neutral quantum definition for generally covariant theories: one can formulate a notion of quantum general covariance in which the perspectives of different observers who use different clocks are related \cite{Vanrietvelde2018, *Hoehn2018, *Hoehn2018_2, Hoehn2019}. The volume time theory in our model does not have a (self-adjoint) Hamiltonian and its solution space is larger compared to the Schr\"odinger time theory; it is not obvious how it can be related to the reduction of a Dirac quantised theory to a specific time coordinate. The authors of \cite{Gotay1984} followed a different route towards quantisation by defining a ``square root'' Hamiltonian generating classical evolution in volume time, which was then subject to a standard Schr\"odinger quantisation with self-adjoint Hamiltonian. It was found that the classical singularity persists also in that quantisation. We plan to investigate this possibility in more detail in further work.

Our cosmological model has another natural  clock, given by the scalar field $\varphi$; indeed free massless scalars are often introduced into quantum cosmology for the main purpose of being used as a clock. The clock $\varphi$ is fast: the classical singularity is at the boundary of its domain $\varphi\rightarrow\pm\infty$. In the simpler case of a model including only this scalar field but no other matter and using $\varphi$ as time, the Wheeler--DeWitt theory does not resolve the singularity; modifying the Hamiltonian and the quantum kinematics using input from loop quantum gravity (LQG) however leads to a loop quantum cosmology with generic singularity resolution (see, e.g., \cite{Ashtekar2011} for details). In our model, Gotay and Demaret's conjecture suggests there should be no singularity resolution in $\varphi$ time. On the other hand, evolution in $\varphi$ is generated by a time-independent Hamiltonian which one would require to be self-adjoint\footnote{We saw earlier that $(v,\varphi)$ can be seen as coordinates on the Rindler wedge, where $\varphi$ is a natural timelike coordinate; the solutions to the Wheeler--DeWitt equation are then closely related to the modes of a scalar field on the Rindler wedge as discussed in, e.g., \cite{Crispino2007}.}, so that the theory should be related to the Schr\"odinger time theory, and the two possibly to the same Dirac quantisation. The latter argument would suggest resolution of the singularity, and thus a counterexample to Gotay and Demaret's conjecture. Again, we leave this question to future work.

\section*{Acknowledgments}


We would like to thank Sean Gryb, Philipp H\"ohn and Jorma Louko for many comments, discussions and helpful insights related to this work, and Martin Bojowald for additional comments on the manuscript. SG would also like to thank Philipp H\"ohn for suggesting the complex semiclassical picture developed in \cite{Bojowald2011}. The work of SG was funded by the Royal Society through a University Research Fellowship (UF160622) and a Research Grant for Research Fellows (RGF\textbackslash R1\textbackslash 180030).

\section*{References}

\bibliographystyle{iopart-num}
\bibliography{bib1_changes}

\end{document}